\documentclass{sigchi}

\CopyrightYear{2019}
\setcopyright{acmcopyright}
\doi{https://doi.org/10.1145/3311350.3347188}
\isbn{978-1-4503-6688-5/19/10}
\conferenceinfo{CHI PLAY '19,}{October 22--25, 2019, Barcelona, Spain}
\acmPrice{\$15.00}

\usepackage{xcolor}
\usepackage{booktabs} 
\usepackage{subfigure}
\usepackage[export]{adjustbox}
\usepackage{balance}       
\usepackage{graphics}      
\usepackage[T1]{fontenc}   
\usepackage{txfonts}
\usepackage{mathptmx}
\usepackage[pdflang={en-US},pdftex]{hyperref}
\usepackage{color}
\usepackage{booktabs}
\usepackage{textcomp}
\usepackage{url}

\usepackage{microtype}        
\usepackage{ccicons}          

\usepackage{todonotes}

\def\plaintitle{SIGCHI Conference Proceedings Format}

\def\emptyauthor{}
\def\plainkeywords{Authors' choice; of terms; separated; by
  semicolons; include commas, within terms only; this section is required.}

\newcommand{\hlinesep}{\unskip\ \vrule\ }

\makeatletter
\def\url@leostyle{%
  \@ifundefined{selectfont}{
    \def\UrlFont{\sf}
  }{
    \def\UrlFont{\small\bf\ttfamily}
  }}
\makeatother
\def\pprw{8.5in}
\def\pprh{11in}

\makeatletter
\def\blfootnote{\xdef\@thefnmark{}\@footnotetext}
\makeatother

\definecolor{linkColor}{RGB}{6,125,233}
\hypersetup{%
  pdftitle={\plaintitle},
  pdfauthor={\emptyauthor},
  pdfkeywords={\plainkeywords},
  pdfdisplaydoctitle=true, 
  bookmarksnumbered,
  pdfstartview={FitH},
  colorlinks,
  citecolor=black,
  filecolor=black,
  linkcolor=black,
  urlcolor=linkColor,
  breaklinks=true,
  hypertexnames=false
}

\begin{document}

\title{Modeling Individual and Team Behavior through Spatio-temporal Analysis}

\numberofauthors{6}
\author{%
  \alignauthor{Sabbir Ahmad\\
    \affaddr{Northeastern University}\\
    \affaddr{Boston, Massachusetts, USA}\\
    \email{ahmad.sab@husky.neu.edu}}\\
  \alignauthor{Andy Bryant\\
    \affaddr{Northeastern University}\\
    \affaddr{Boston, Massachusetts, USA}\\
    \email{A@andymbryant.com}}\\
  \alignauthor{Erica Kleinman\\
    \affaddr{Northeastern University}\\
    \affaddr{Boston, Massachusetts, USA}\\
    \email{kleinman.e@husky.neu.edu}}\\
  \alignauthor{Zhaoqing Teng\\
    \affaddr{Northeastern University}\\
    \affaddr{Boston, Massachusetts, USA}\\
    \email{teng.z@husky.neu.edu}}\\
  \alignauthor{Truong-Huy D. Nguyen$^\dagger$\\
    \affaddr{Fordham University}\\
    \affaddr{Bronx, New York, USA}\\
    \email{truonghuy@gmail.com}}\\
  \alignauthor{Magy Seif El-Nasr\\
    \affaddr{Northeastern University}\\
    \affaddr{Boston, Massachusetts, USA}\\
    \email{magy@northeastern.edu}}\\
}

\maketitle

\begin{abstract}
Modeling players' behaviors in games has gained increased momentum in the past few years. This area of research has wide applications, including modeling learners and understanding player strategies, to mention a few. In this paper, we present a new methodology, called \textit{Interactive Behavior Analytics (IBA)}, comprised of two visualization systems, a labeling mechanism, and abstraction algorithms that use Dynamic Time Warping and clustering algorithms. The methodology is packaged in a seamless interface to facilitate knowledge discovery from game data. We demonstrate the use of this methodology with data from two multiplayer team-based games: \textit{BoomTown}, a game developed by Gallup, and \textit{DotA 2}. The results of this work show the effectiveness of this method in modeling, and developing human-interpretable models of team and individual behavior.

\end{abstract}




\section{Introduction}\label{sec:introduction}
\blfootnote{$^\dagger$recently Huy has taken a job as a software engineer at Google, YouTube team}
Analysis of player behavior, specifically to understand individual or team strategies, is an important research focus due to its overarching utility in game development. Understanding individual or team problem solving behaviors and strategies can benefit the development of personalized games, Non-Player Characters (NPCs), AI agents, and game guides. This can also aid in assessment and the development of scaffolding techniques to improve applications, such as health or learning games. Additionally, understanding team strategies is critical for E-Sports games as it can be used to facilitate team training, play analysis, and game balancing. Such analysis, however, is highly complex due to the dynamic nature of modern games. This is especially true in multiplayer team-based games, in which teamwork affects individual players' behaviors and strategies, and subsequently adds to the complexity of these game environments. 

Fortunately, there is now an abundance of data available, which opens new opportunities for this work. In the past few years, several researchers proposed various techniques. Some used machine learning and statistical analysis to predict game outcomes~\cite{agarwala2014learning, semenov2016performance, ong2015player} and model individual and team tactics or strategies~\cite{weber2009data, cavadenti2016did, yang2014identifying, mahlmann2016esports}. Other works investigated cluster analysis to group players based on their spatio-temporal behaviors~\cite{drachen2012guns, ramirez2010player, nascimento2017profiling, neidhardt2015team, drachen2014skill}. Thus far, most of the methods used by previous work use machine learning algorithms or inferential statistics. Unfortunately, most machine learning techniques result in models that are hard to interpret or adjust by human experts. A few studies have explored the development of new methods~\cite{weber2009data, yang2014identifying, mahlmann2016esports, drachen2014skill}, but such methods are game specific, and thus do not generalize. Therefore, there is a dire need for new methodological approaches that are generalizable across games and that produce human-interpretable models. This paper attempts to address this need. 

We propose \textit{Interactive Behavior Analytics (IBA)} as a methodological approach to study player behaviors across various types of games. The methodology consists of two visualization systems: \textit{StratMapper} (developed based on \textit{G-Player}~\cite{canossa2016g}) and \textit{Glyph}~\cite{nguyen2015glyph}. \textit{StratMapper} presents a spatio-temporal visualization of game events, which is used to analyze players and team behaviors throughout a game or a game level. It includes a built-in labeling mechanism, enabling a human user to identify high level abstractions over low-level data, and therefore label or identify tactical and strategic decisions as well as important high level coordination or correction behaviors. This labeled data is then visualized by \textit{Glyph} to show player and team behaviors in a sequential manner. When behaviors are visualized in \textit{Glyph}, users can better see if the labels captured the essence of the strategies, tactics, or high level behaviors investigated. They then can re-evaluate their labels and adjust them in \textit{StratMapper}. This type of iterative process is not possible with current tools and is important for developing abstractions that can turn low-level data into knowledge. To test this approach, we apply it to two games: \textit{BoomTown}, a multiplayer online game developed by Gallup, and \textit{DotA 2} (Valve Corporation, 2013), a popular multiplayer online battle arena game. Inter-rater reliability measures were used to test reliability of the labels developed. It should be noted that both \textit{Glyph} and \textit{StratMapper} have been tested for usability before integration into \textit{IBA}, and thus we will not discuss their evaluation as visualization systems in this paper. \textit{IBA} allowed us to develop human-interpretable models of players' plans and strategies, uncovering critical insights about players' decisions and behaviors. Thus, we propose \textit{IBA} as a generalizable method to study individual and team behavior and produce interpretable models of strategies and tactics.

\textit{IBA} is targeted towards researchers, game designers, and developers, as it gives them a way to visualize players' behaviors. It facilitates a qualitative analysis approach to data in order to identify behaviors, tactics, and strategies that players devised through their interaction. It also highlights errors and cyclic behaviors allowing designers to find imbalance in game mechanics, adjust the game to the target audience, or develop scaffolding or adaptive methods. \textit{IBA} is a general framework where different visualization systems can be used. In place of \textit{Glyph}, for example, other visualization systems presented in~\cite{thawonmas2008visualization, andersen2010gameplay, wallner2012spatiotemporal,osborn2018visualizing} can be used, since these systems use similar graph-link visualizations to represent player activities and game states.

The rest of the paper is structured as follows. 
First, we discuss related works. We then discuss \textit{IBA} in detail. Then, we discuss the process and results of applying the proposed methodology to \textit{BoomTown} and \textit{DotA 2}, respectively. Finally, we conclude the paper with a discussion and future work.

\section{Related Works}\label{sec:related_works}

Understanding player strategies in games has been a relevant research topic in the past few years. Several survey papers discussed player modeling methods, including action modeling, strategy modeling, and profiling, e.g.,~\cite{yannakakis2013player, bakkes2012player, smith2011inclusive}. We will focus the discussion on behavior modeling using game data.

Related to understanding players' strategies, plan recognition has been proposed as a technique to uncover players' plans and goals. Most plan recognition techniques use an inference algorithm to infer plans/goals given a set of actions and a plan library developed by an expert~\cite{sukthankar2014plan}. Fagan and Cunningham~\cite{fagan2003case} used case-based plan recognition to model players using data from \textit{Space Invaders} (Taito, 1978), where plan libraries were developed by observing gameplay. Constructing such plan libraries for simple games is feasible, however, current Real-Time Strategy (RTS) and multiplayer games are far too complex for such techniques. Several researchers have explored other approaches to infer tactics and strategies, such as using inverse reinforcement learning, Monte Carlo methods, genetic algorithms, and case-based reasoning~\cite{ontanon2008learning, balla2009uct, ponsen2005automatically, aha2005learning}. These techniques, however, do not capture the complexity in current games. 

Additionally, previous works proposed various machine learning approaches to predict game outcome depending on disparate factors, including game statistics or team composition~\cite{agarwala2014learning, rioult2014mining, semenov2016performance, yang2013extracting, ong2015player, pobiedina2013successful, summerville2016draft}. For example, Sapienza et al.~\cite{sapienza2018non} used decision trees to identify features that were meaningful for performance prediction. Similarly,~\cite{miller2010group, sapienza2018individual} used spatio-temporal data and player actions to model player behavior. Further, Mahlmann et al.~\cite{mahlmann2010predicting} used aggregated metrics of in-game data to predict churn. However, due to the nature of the metrics used in this work, it was hard to decipher the specifics of obstacles that led players to quit, which limits the use of these models for design adjustment. 

Some exploratory works used clustering to group players based on their roles and behaviors to understand individual differences. For example, Drachen et al.~\cite{drachen2012guns} used Simplex Volume Maximization and k-means clustering with qualitative analysis of \textit{Tera} (Bluehole Studio, 2011) and \textit{Battlefield 2} (Digital Illusions CE, 2005). Ramirez et al.~\cite{ramirez2010player} proposed a meta-classification approach based on roles, preferred areas, and social interactions using data from a first-person social hunting game. Nascimento et al.~\cite{nascimento2017profiling} investigated clustering players of \textit{League of Legends} (Riot Games, 2009) based on team performance metrics, team behavior patterns, and successful/unsuccessful team behavior profiles. Although these works are insightful, they focus on grouping players based on their behaviors, and thus fail to capture the dynamics of strategic decision-making given context and environmental changes in a human-interpretable way.  

Few works have investigated strategic gameplay and decision-making. For example, Weber et al.~\cite{weber2009data} applied classification and regression methods using labeled features to classify and predict strategic decisions regarding unit or building types in \textit{StarCraft}'s (Blizzard Entertainment, 1998) one-versus-one matches. They used expert knowledge to generate rule-based models; the approach was limited to \textit{StarCraft} data. Cavadenti et al.~\cite{cavadenti2016did} used discriminant pattern mining to identify strategic patterns (e.g., frequent items used by the players) in \textit{DotA 2}. 

While most of the previous works focused on individual players, some have investigated team behavior and strategy. Yang et al.~\cite{yang2014identifying} proposed a data driven approach to identify combat patterns in \textit{DotA 2} by generating rules that identify outline sequences of combat actions that lead to victory. On the other hand, Mahlmann et al.~\cite{mahlmann2016esports} presented performance evaluation metrics to generate win predictions in \textit{DotA 2}. Drachen et al.~\cite{drachen2014skill} extracted spatio-temporal measures of zone changes and team distributions (distances between heroes on a team) in \textit{DotA 2}, and showed differences in positioning and variance between professional and amateur players. 

While there is much work in this area, there are several general limitations to previous approaches: (a) they use aggregate data, and thus, do not conserve sequences of behaviors or actions to investigate the dynamics of the decision-making process, (b) with the exception of decision trees and Bayesian Networks, models produced are not human-interpretable or correctable, (c) most of the labeling methods used are game specific, and thus are not generalizable, and (d) they focus on prediction or clustering rather than understanding players' decision-making processes. To address these issues, we present an alternative approach to player behavior modeling --- an approach that is similar to Pfister's work on human-centered machine learning, which aims to create human-interpretable machine learning models using visualization~\cite{strobelt2018lstmvis, strobelt2019s}. In the spirit of that approach, we use several visualization systems and a human in the loop architecture to unpack game behavior, developing human-interpretable models of strategies, tactics, and the decision-making processes.

\begin{figure}
    \includegraphics[width=\columnwidth]{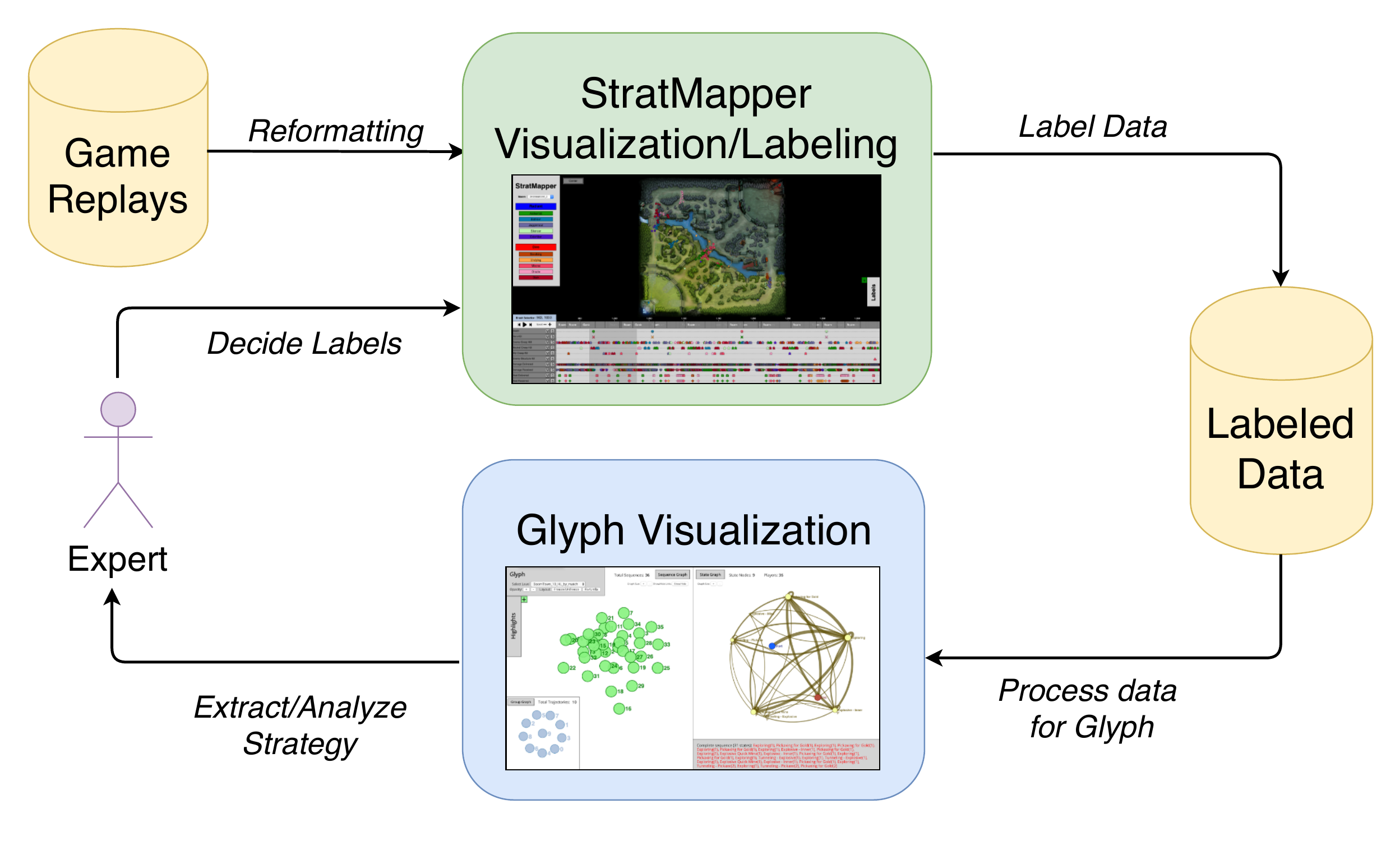}
    \caption{Overall approach of Interactive Behavior Analytics method.}
    \label{fig:overall_method}
\end{figure}

\section{Interactive Behavior Analytics Method}\label{sec:methodology}
To develop and model strategies and tactics in an interpretable way, we propose a new methodology, called \textit{Interactive Behavior Analytics (IBA)}, shown in Figure~\ref{fig:overall_method}. \textit{IBA} is composed of multiple systems, including two visualization systems, a labeling interface, and several data processing algorithms. \textit{IBA}'s architecture integrates a human-in-the-loop method, where human intervention is used to develop abstractions, labels, and ways to interpret the data. Below we detail \textit{IBA}, outlining its various tools and processes.

\subsection{Tools}\label{subsec:tools}

\subsubsection{StratMapper}\label{subsubsec:StratMapper}
\begin{figure}
    \includegraphics[width=\columnwidth]{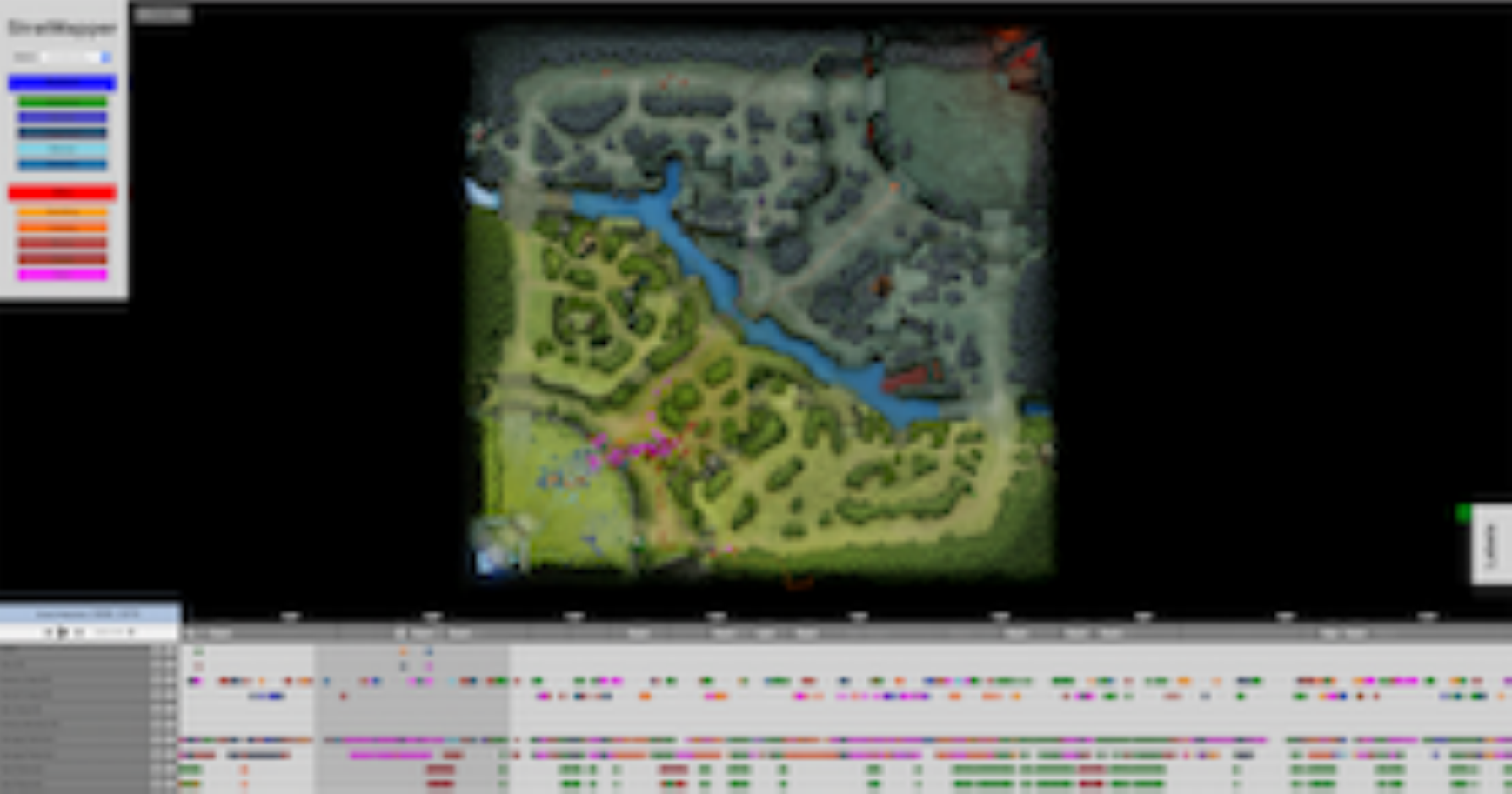}
    \caption{\textit{StratMapper} interface. The two-dimensional map appears in the center. The units and unit teams are listed in the upper left, and the timeline is on the bottom, showing a chronological series of events.}
    \label{fig:sm_interface}
\end{figure}

\textit{StratMapper} is a visualization and labeling system for game data. It offers an abstract playback of events, and allows users to filter data and zoom into behaviors of interest. Figure~\ref{fig:sm_interface} shows a screenshot of the interface. At the center of the application there is a two-dimensional game map, which can be zoomed in and out. Below it is a timeline, which can also be zoomed in and out, visualizing all events in the game. Each event icon in the timeline corresponds to an icon on the game map, as well as an event type on the $y$-axis displayed with the color assigned to the corresponding unit. A tool-tip appears with detailed information upon hovering over the corresponding event icon.

The timeline allows users to select a temporal period of interest by placing start and end markers, which would then filter events and users on the game map, and display only those that appear during the selected period. Event types, listed on the left of the timeline, as well as units and unit teams, listed in the top left of the interface, can also be filtered. In addition, using \textit{StratMapper} to select several events on the timeline and units involved, one can compose a higher-order event from event primitives. We developed a labeling component to the interface which allows users to label such higher-order events. This then saves the selected events, time window, and units, under the specified label (discussed later). Such saved labels are shared among all collaborators using the visualization and can be retrieved, deleted, or corrected. These labels, along with the data, can also be exported for further processing.

\subsubsection{Glyph Interface and Underlying Structure}\label{subsubsec:glyph_interface}
\begin{figure}
    \includegraphics[width=\columnwidth]{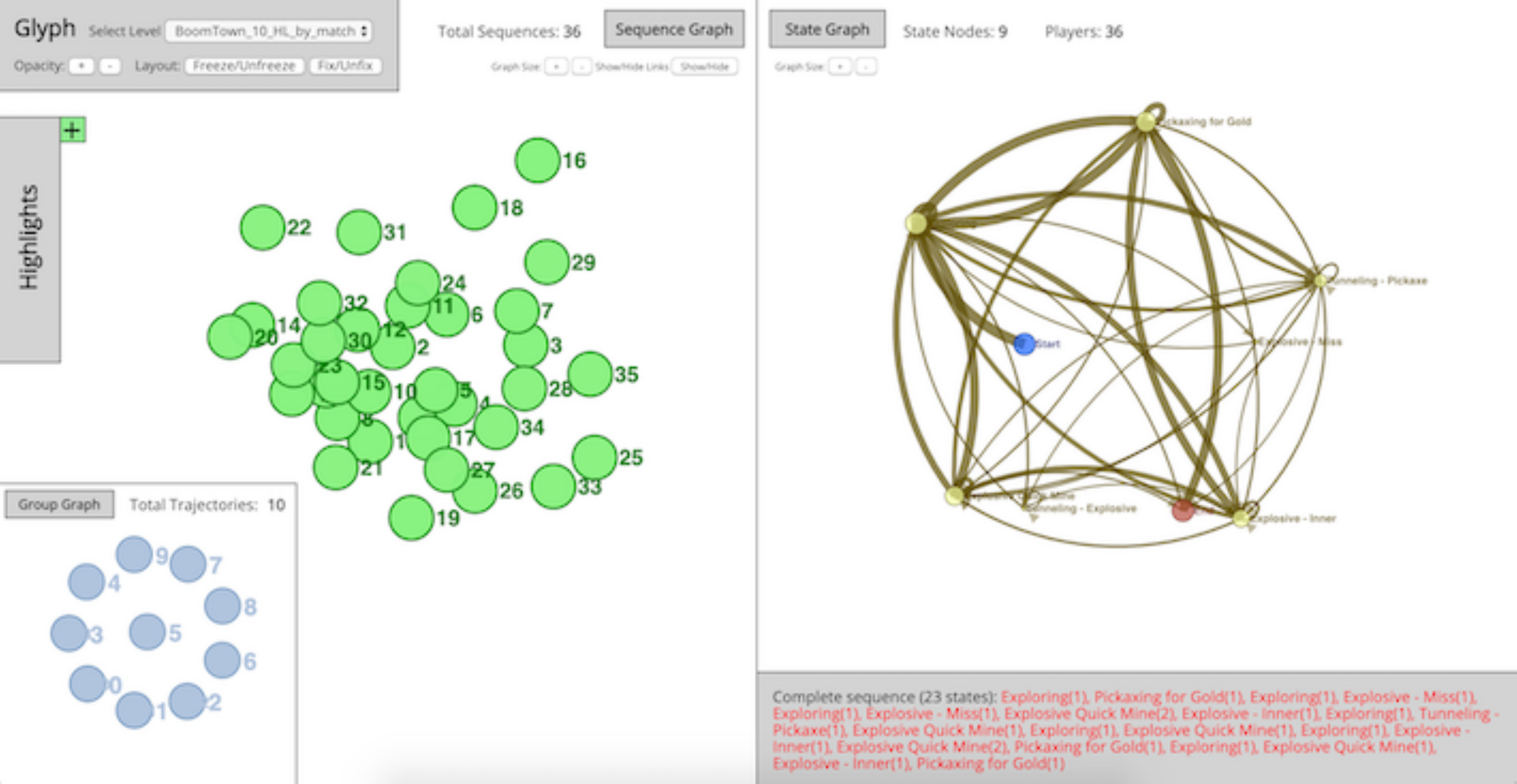}
    \caption{\textit{Glyph} interface showing the three visual windows: a state graph (right), a sequence graph (left), and a group graph (bottom left) with filters and user controls in the upper-left.}
    \label{fig:glyph_interface}
\end{figure}

\textit{Glyph}~\cite{nguyen2015glyph} is a visualization system that allows easy understanding as well as investigation of player strategy~\cite{Nguyen2015visual, nguyen2015interactive}. While it was originally designed to investigate individual behavior, we extended it to incorporate team behaviors. The new version, shown in Figure~\ref{fig:glyph_interface}, is composed of three visual representations (as opposed to two in the original system): a state graph (right), a sequence graph (upper left) and a group graph (bottom left). The state graph shows action paths as a node-link diagram, and the sequence graph encodes action sequences as nodes. These graphs were discussed in~\cite{nguyen2015glyph}. The group graph, a new window to the \textit{Glyph} system, shows nodes for team behaviors. We describe these below. 

\textbf{The State Graph} displays a node-link graph where game states are represented as nodes and player actions as links~\cite{nguyen2015glyph}. The node size and link thickness represent the number of players visiting the states and actions. 

In \textbf{The Sequence Graph} each node represents a full sequence of state-actions. The distance between the nodes depends on the similarity between the sequence patterns: the more similar the sequence, the closer the nodes are in the sequence graph. This enables the system to cluster similar nodes together, where similarity is calculated using Dynamic Time Warping (DTW)~\cite{berndt1994using}. 

To calculate similarity using DTW, we need to define a way to compare two game states, i.e., state difference. State difference can vary depending on the game type and how the data is abstracted. This can be defined by game designers. Using this measure, the DTW algorithm can be used to deduce dissimilarity between two sequences. 

\textbf{The Group Graph} is a new addition to~\cite{nguyen2015glyph}. It consists of nodes representing a full sequence of team behaviors (game state and actions) instead of individual behaviors, like in the sequence graph.

\textbf{The complete sequence of states} is shown below the state graph upon selecting a node in the sequence or the group graph. Note that the number associated with each state inside the parenthesis represents the number of times the state re-occurred in that position of the sequence.

\subsection{IBA}\label{subsec:apply_method}
\textit{IBA} is composed of two main processes (as shown in Figure \ref{fig:overall_method}): (a) developing, identifying, and applying labels, and (b) visualizing labeled data using \textit{Glyph} or other visualization systems. Human intervention and correction happens across these two phases, as we will discuss below. 

\subsubsection{Labeling Process}\label{subsubsec:labeling_process}

\begin{figure}
    \includegraphics[width=\columnwidth]{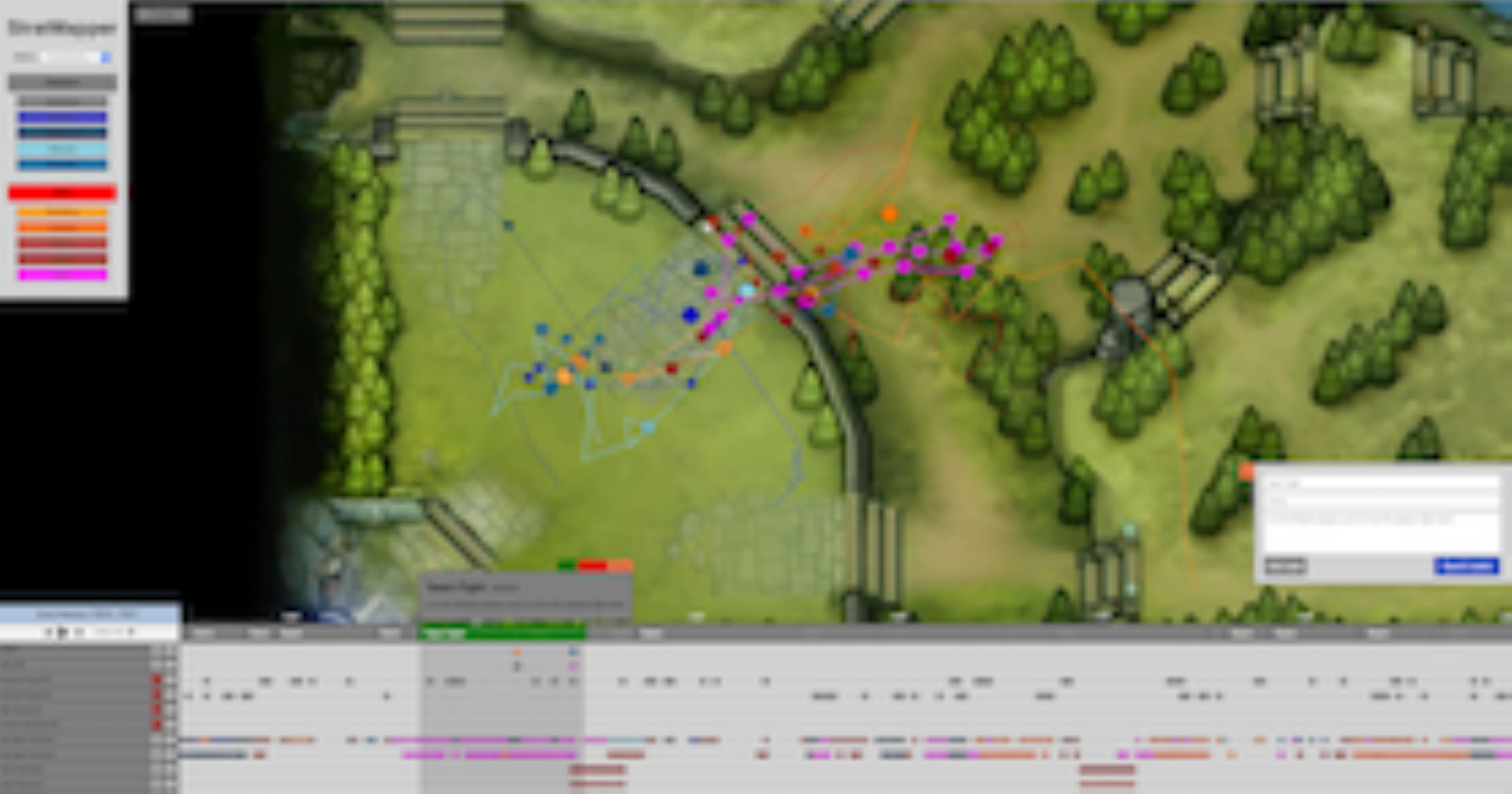}
    \caption{\textit{StratMapper} interface showing a selected label on the timeline and the corresponding units and events on the map. }
    \label{fig:sm_label1}
\end{figure}

Developing labels is similar to developing codes for qualitative thematic analysis. The process is iterative. It begins with members of the research team familiarizing themselves with the game to be labeled. In the case of \textit{BoomTown}, this involved playing several rounds to get comfortable with the gameplay, goals, and strategies available. For \textit{DotA 2}, this involved an in depth analysis of online \textit{DotA} community resources, including forums, tutorials, and professional gameplay streams. In both cases, once the researchers are familiar and comfortable with the game's action space, they then discuss the types of strategic actions taken during gameplay. Using this discussion, researchers then adopt a thematic analysis approach to identify re-occurring patterns and themes within the game data visualized in \textit{StratMapper}. From there, they then identify a set of labels for decision-making, strategic, and tactical behaviors.

Once the labels are identified, team members individually label the game using \textit{StratMapper}. To apply a label in \textit{StratMapper}, the user selects all units and events relevant to a behavior in a specific time period. This is done by muting irrelevant units and events, and using the brush tool to select a particular time period in the timeline (see Figure~\ref{fig:sm_label1}). An Inter-Rater Reliability (IRR) measure, such as Cohen's Kappa measure \cite{bordens2002research}, is then computed as a measure of reliability of the labels produced. The research team may choose to discuss the labels if the reliability is very low, in order to identify and refine labels with high variance or conceptual issues.

Once this process is complete, the labels go through another process of refinement where the labeled data are visualized in \textit{Glyph} (discussed below). Using \textit{Glyph}, researchers can make sense of the labels as a sequence of behaviors and patterns to solve particular game levels. Sometimes the labels may be incomplete or the resulting patterns may be incomprehensible. Thus, the team discusses their interpretations of the resulting sequences, and may iterate and revise the labels accordingly. This process is generalizable and can be done for different games, as we shall see in the case studies below.

\begin{figure}
    \includegraphics[width=\columnwidth]{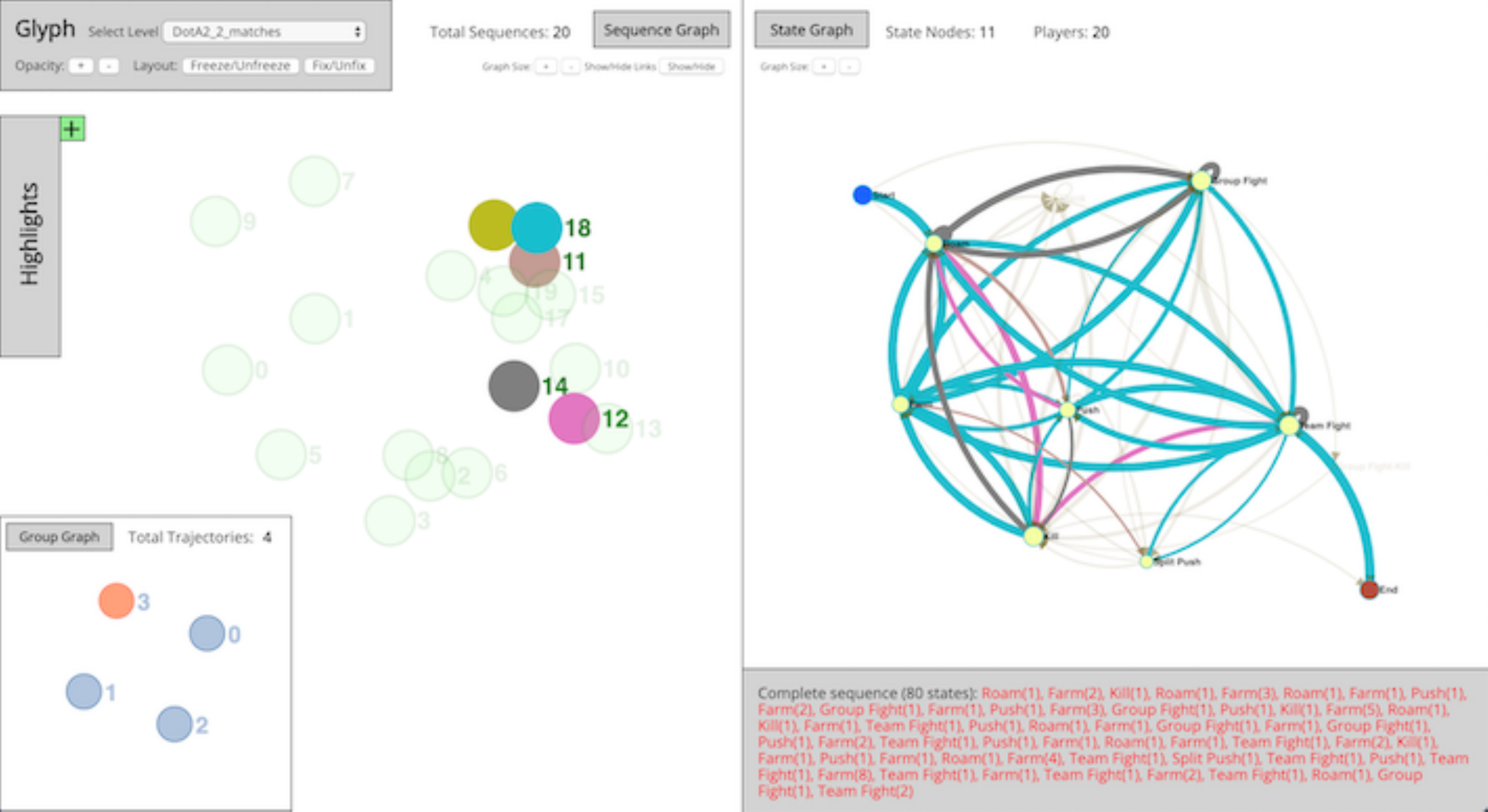}
    \caption{Selecting team 3 in the group graph (bottom left) highlights the team members (11, 12, 14, 16, 18) and their trajectories in the sequence graph (left) and state graph (right).}
    \label{fig:group_graph}
\end{figure}

\subsubsection{Visualization of Labeled Data}\label{subsubsec:viz_labeled_data}
As discussed above, after a set of labels are applied through \textit{StratMapper}, the labels, and associated information are exported and processed to convert the labels to the input format of \textit{Glyph}. In the conversion process, the states and links of the state graph in \textit{Glyph} are determined. We consider each label as a different state of the game visualized as a node in the state graph. The directed link between the two states (labels) demonstrates the change of player behavior. Individual player behavior is defined as a vector of labels sorted by time. 
Team behavior is, similarly, a vector of labels. We consider any label that includes more than one player from the same team as a team behavior. These labels, as a sequence, represent the team behavior. Note that all team behavior labels are also considered individual behaviors to enable visualization in the individual player trajectory. 

However, there are two challenges with this process. First, labels can overlap over time for a given team. For example, in \textit{DotA 2}, three players are engaged in a ``Team Fight'', while another two are engaged in a ``Gank'' simultaneously. This would result in two separate labels overlapping over time. To resolve this issue, we introduce a separate node as a game state in the state graph, which is named with both the labels; for this example, the node would be named as ``Team Fight - Gank''. This technique is applied whenever we encounter two or more overlapping labels. Second, in order for us to visualize both individual and team behaviors, we introduced the group graph as discussed previously. Similar to an individual player trajectory, the team trajectory is visualized by its own node-link diagram in the state graph, and by a unique node in the group graph, displayed next to the sequence graph. When a node in the group graph is selected, the corresponding nodes of the team members in the sequence graph, and their trajectories in the state graph, are highlighted (see Figure~\ref{fig:group_graph}).




\section{BoomTown Case Study}
\label{subsec:apply_boomtown}
In this section, we discuss how we used \textit{IBA} to analyze player behaviors in \textit{BoomTown} --- a multiplayer online team-based game. The game is played in teams of 3-6, where players can use several items to clear rocks and mine gold within a two-dimensional map, in order to pursue the team goal of maximizing the amount of gold collected. The game consists of six rounds; at the beginning of each round the team elects a team leader and identifies a set of explosives to use during the round. The team score is determined by the total amount of gold earned within the game. Our research goal here was to identify successful strategies, and thus we used \textit{IBA} to identify and compare strategies of low vs. high performing teams.   

\subsection{Process}
\textbf{Labels.} As discussed above, the researchers identified labels for \textit{BoomTown} after playing the game for several rounds to become familiar with the mechanics, play space, and types of actions one can take. Then, the researchers developed an initial list of labels consisting of four labels: ``Tunneling - Optimize'', players tunneled into the center of a rock formation before detonating an explosive; ``Tunneling - Passage'', players tunneled through a rock formation to get to a different part of the map; ``Pickaxing for Gold'', players spent time using the pickaxe on rocks; and ``Exploring'', players moved around the map but did nothing. 

This set was then used to label three games for \textit{Glyph} visualization. Upon review of the visualization, the researchers decided that the initial set of labels was not comprehensive, i.e., did not account for the various ways in which players used explosive items. Thus, the labels were revised and a final set of labels was produced consisting of seven labels: ``Pickaxing for Gold''; ``Exploring''; ``Tunneling - Pickaxe'', the player uses a pickaxe to get to another area or clear an area for a mine; ``Tunneling - Explosive'', a player uses an explosive to get to another area or clear an area for a mine; ``Explosive - Inner'', a player uses a pickaxe to optimize an item's yield; ``Explosive - Miss'', a player uses an explosive away from a rock formation; ``Explosive - Quick Mine'', a player uses an explosive on the outer boundary of a rocky area to quickly obtain gold. In the final iteration, researchers used the labels to label ten games, and concluded that those labels were comprehensive and had captured observed strategic and tactical behaviors. Inter-rater reliability was performed and the result showed acceptable reliability. 


\textbf{Data.}
We collected \textit{BoomTown} replay files from Gallup. First, we investigated the data of 100 high-performing and 100 low-performing teams. This initial investigation with 200 teams was done to allow us to evaluate the population and thus select teams for our labeling that mirror the population. We then selected 10 games to label with five high-performing and five low-performing teams. A high-performing team is defined by the highest average amount of gold collected per person in a team, and the opposite for low-performing. Investigating the teams, we see that there is a bias towards replays, where people who replayed the game tend not to be low performers. Specifically, out of the 100 high performers, 52 were replays and out of the 100 worst performers, only 7 were replays. Thus, our selected groups for labeling mirrored that bias.

The data for 10 games was labeled following the procedure discussed above. The visualization in \textit{Glyph} is detailed in the following section. It took approximately 6 hours for each researcher to apply the \textit{BoomTown} labels. The IRR score, using Cohen's Kappa, was computed for the labels, resulting in .95 with 96\% agreement, which was deemed acceptable. 



\begin{figure}%
    \centering
    \subfigure[]{
    \label{fig:boomtown_overall_seq}%
    \includegraphics[width=0.42\columnwidth]{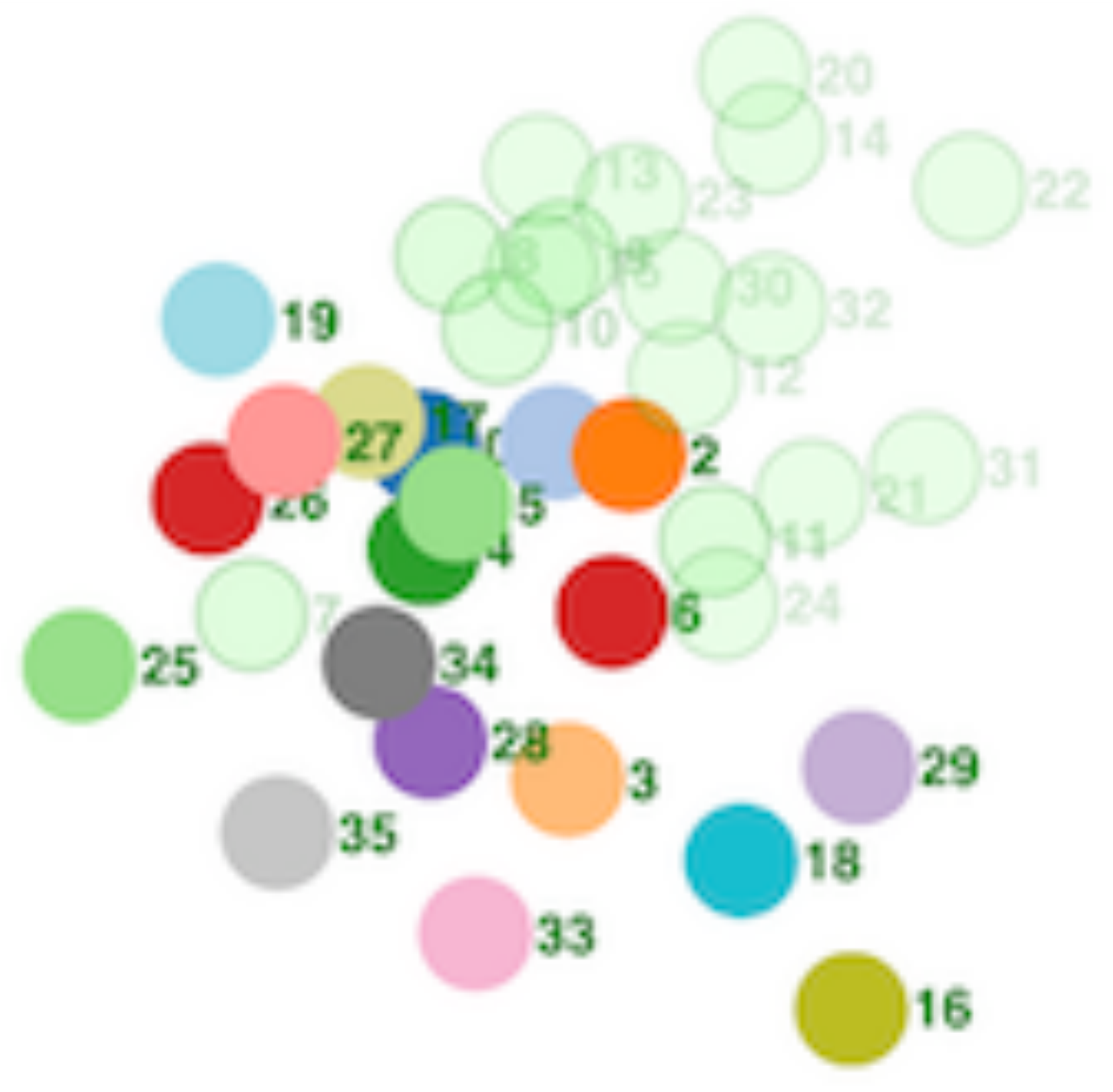}}
    \hlinesep
    \subfigure[]{
    \label{fig:boomtown_twoteam_seq}%
    \includegraphics[width=0.42\columnwidth]{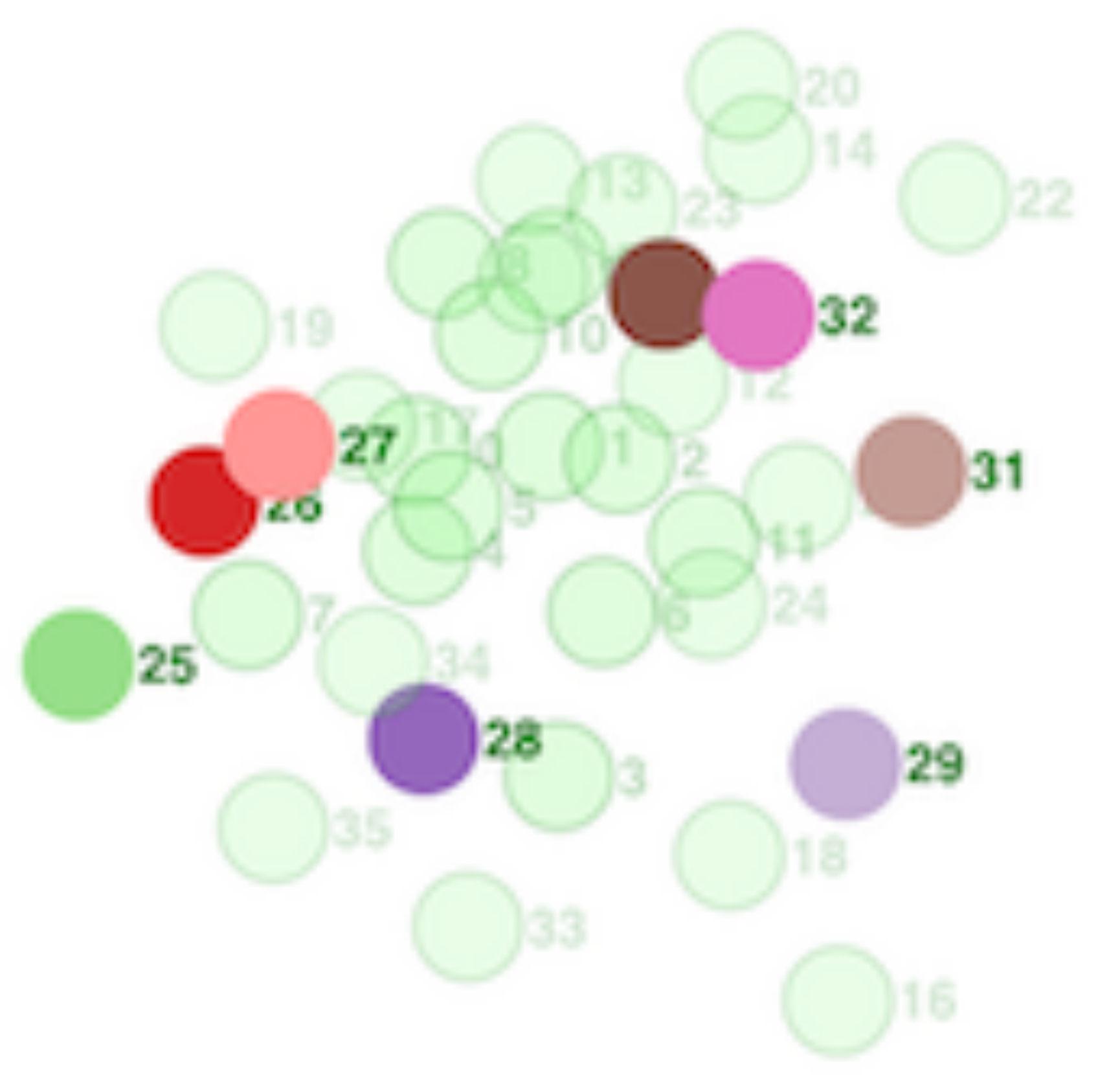}}
    \caption{Sequence graph showing (a) high-performing team players in colored nodes and (b) players from one high-performing (25, 26, 27, 28, 29) and one low-performing (30, 31, 32) team.}
    \label{fig:boomtown_seq_graph}
\end{figure}

\subsection{Results}\label{subsec:result_boomtown}
With \textit{Glyph}, it is possible to see differences in the gameplay and behavior sequences of high-performing and low-performing teams, allowing us to draw conclusions regarding their strategic approach. We divide our analysis into two phases: Investigation of Overall Performance and Investigation of Team Behaviors Per Round.

\subsubsection{Investigation of Overall Performance}\label{subsubsec:boomtown_overall}
 
Using the sequence graph from \textit{Glyph}, shown in Figure~\ref{fig:boomtown_seq_graph}, we highlighted the high-performance teams in~\ref{fig:boomtown_seq_graph}(a), and highlighted one high and one low-performance team in~\ref{fig:boomtown_seq_graph}(b). From this, we can see that the player trajectories of high-performing teams are clearly different from the player trajectories of the low-performing teams. 
It should be noted that the sequence graph uses Dynamic Time Warping (DTW), as discussed in the previous section. A clustering mechanism is used to lay out the nodes using the distances computed via DTW. As can be seen from Figure~\ref{fig:boomtown_overall_seq}, the high-performing team nodes are clustered together (colored nodes) and are more distant from the low-performing team nodes, which are clustered together as well. This linear boundary between the two groups shows that there is a clear difference between the two groups' sequences. We investigated further to see if we can understand the difference between the groups on a more qualitative level. 

\subsubsection{Investigation of Team Behaviors Per Round}\label{subsubsec:boomtown_per_round}
We visualized the labeled data based on round to see and compare the activities of players during different rounds. Using \textit{Glyph} we uncovered some patterns that differentiated the high-performance teams from low-performance teams. These patterns include: frequent use of optimized and efficient explosive placement, and doing more in the game than others. To illustrate these patterns, we discuss a comparison between a single low-performing team (nodes: 30, 31, 32) and a high-performing team (nodes: 25, 26, 27, 28, 29). Looking at the respective sequence graphs (Figure~\ref{fig:boomtown_twoteam_seq}) in \textit{Glyph}, we can see that the players' sequences are quite different from each other.

\begin{figure}%
    \centering
    \subfigure[]{
    \label{fig:boomtown_round2_high}%
    \includegraphics[width=0.48\columnwidth]{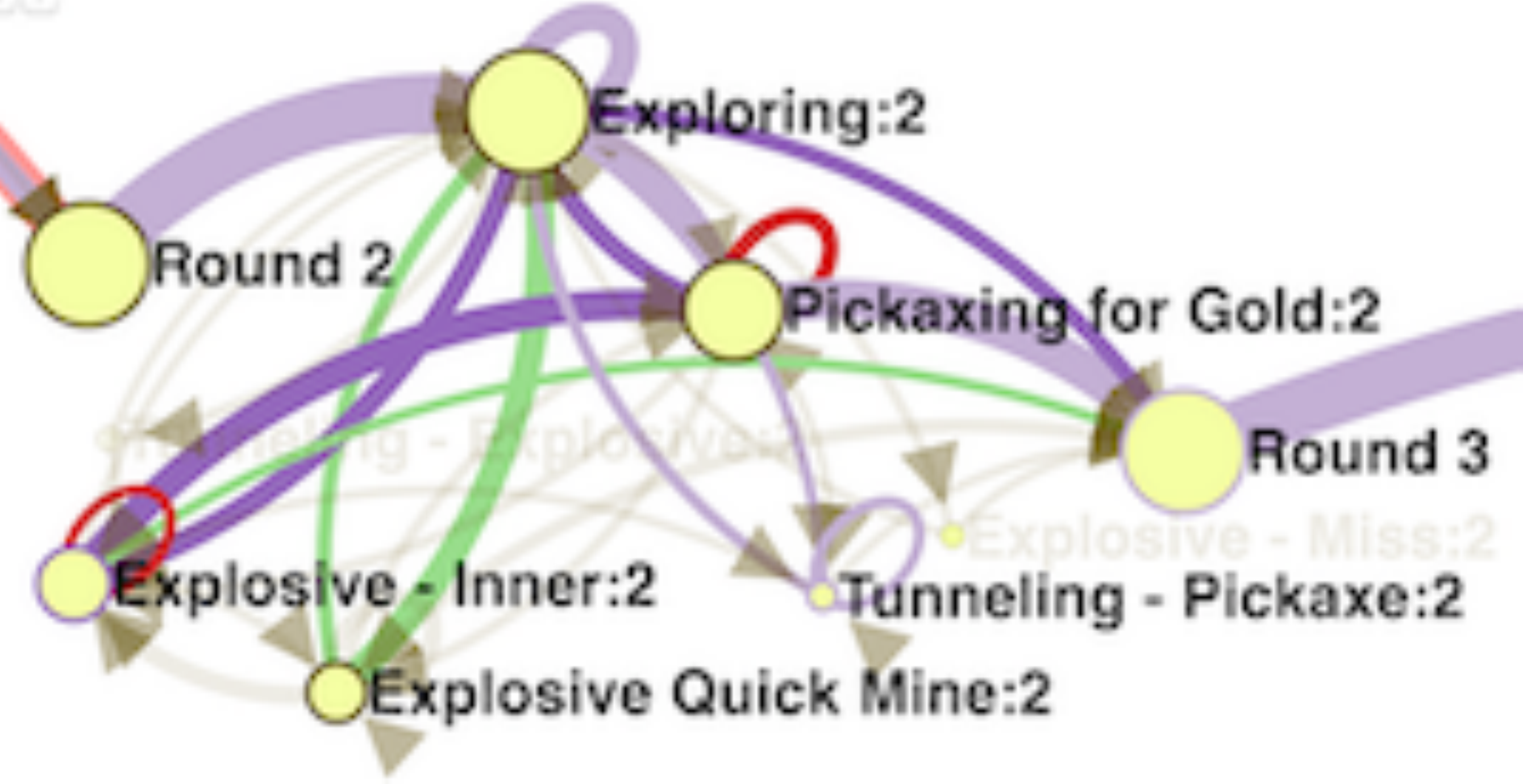}}
    \subfigure[]{
    \label{fig:boomtown_round2_low}%
    \includegraphics[width=0.48\columnwidth]{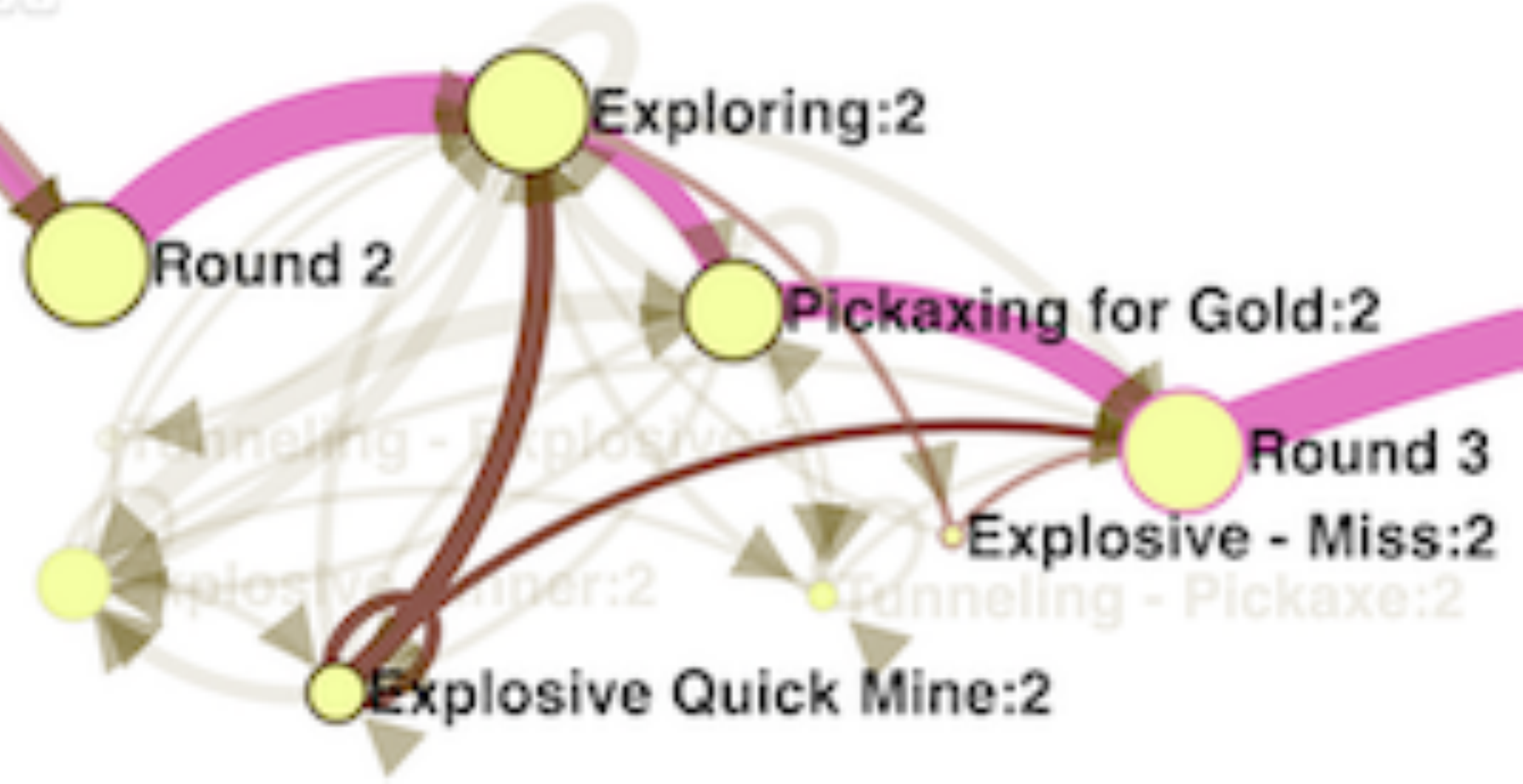}}
    \caption{State sequences in Round 2 for (a) the high-performing team and (b) the low-performing team in the state graph.}
    \label{fig:boomtown_round2}
\end{figure}

Investigating the patterns by round allows us to see the strategic differences between the teams. Due to space constraints, we only discuss the analysis for one round in this paper. The differences are apparent in round 2 (Figure~\ref{fig:boomtown_round2}), where the high-performing team utilized a strategy of optimizing explosives by setting them off within rock formations in order to obtain more gold, as seen by them visiting the ``Explosive - Inner'' state. They also have more transitions in and out of ``Pickaxing for Gold'' than the low-performing teams. By contrast, the low-performing team visited the state ``Explosive - Miss'', indicating that a player used an explosive in an area with no rocks and no gold. These findings indicate that high-performing teams engaged in more focused behaviors, as they utilized the new explosive items to efficiently mine and further pursue their goal. The low-performing team appears to be testing the new items, but not in a strategically focused manner.


\begin{figure}
    \includegraphics[width=\columnwidth]{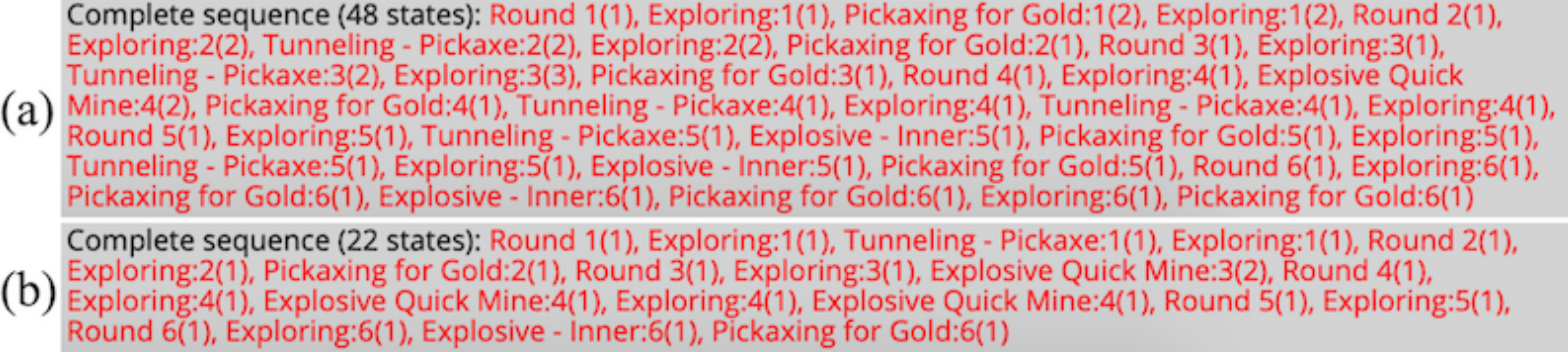}
    \caption{State sequences of two players, (a) one from a high-performing team, with 48 states, and (b) one from a low-performing team, with 22 states.}
    \label{fig:boomtown_com_seq}
\end{figure}


Further, when investigating the complete state sequences displayed per player within \textit{Glyph}, as shown in Figure~\ref{fig:boomtown_com_seq} for two players, we see that the high-performing team player had more state transitions per round than the low-performing team player, indicating that players on high performing teams were more active, visited more states, and displayed more behaviors in the given time. By comparison, the low-performing team member appears to have spent more time wandering the map or doing nothing rather than performing goal-oriented actions. Inspecting other players in high and low-performing teams, we observed this phenomenon occurring independent of the number of players on the team.


\section{DotA 2 Case Study}
\label{subsec:apply_dota2}

\textit{DotA 2} is a popular multiplayer online battle arena (MOBA) game, and a sequel to \textit{Defense of the Ancients (DotA)}. Each match is played between two five-player teams, and each team has a base located in the opposite corners of the map from each other. Each player controls a character called a ``hero'', and collects experience points and items during gameplay to defeat the opposing team's heroes. The goal is to destroy a large structure located in the opposing team's base known as the ``Ancient''. The first team to do so wins the game.

\subsection{Process}

\textbf{Labels. }For \textit{DotA 2}, two researchers, both with previous experience playing MOBA games, performed an analysis of online \textit{DotA 2} forums, tutorials, and gameplay commentaries to familiarize themselves with existing strategic behaviors as recognized by the gameplay community. From this analysis, the researchers used an iterative process to develop a set of twenty-four labels that they believed captured relevant behaviors of strategic gameplay in high level \textit{DotA 2} matches. 
For the purpose of this paper, this set was reduced to nine labels via the omission of highly specific labels and the collapse of similar labels. These labels are: ``Push'', representing the event that one team destroys the enemy's structures, including towers, barracks, and ancients, as well as any behavior identifiable as an attempt to push a lane towards the enemy side of the map, or push back enemy entities from one's base; ``Farm'', meaning the process by which a hero gains experience and gold to become stronger through the continuous killing of NPC entities; ``Kill'', which shows that a hero used abilities or normal attacks to kill an enemy; ``Group Fight'', representing a small fight between two teams with only part of the team involved; ``Team Fight'', representing a fight between two teams that most of both teams were involved in; ``Roam'', which represents a behavior in which heroes move about the map; ``Gank'', representing unexpected heroes arriving in a lane to attack and hopefully kill the enemy; ``Roshan'', indicating that players engaged the NPC entity named Roshan in combat; and finally, ``Split Push'', a strategy in which some of the team is pushing one lane while others are distracting the enemy.

\begin{figure}%
    \centering
    \subfigure[]{
    \label{fig:Radiant1}%
    \includegraphics[width=0.40\columnwidth]{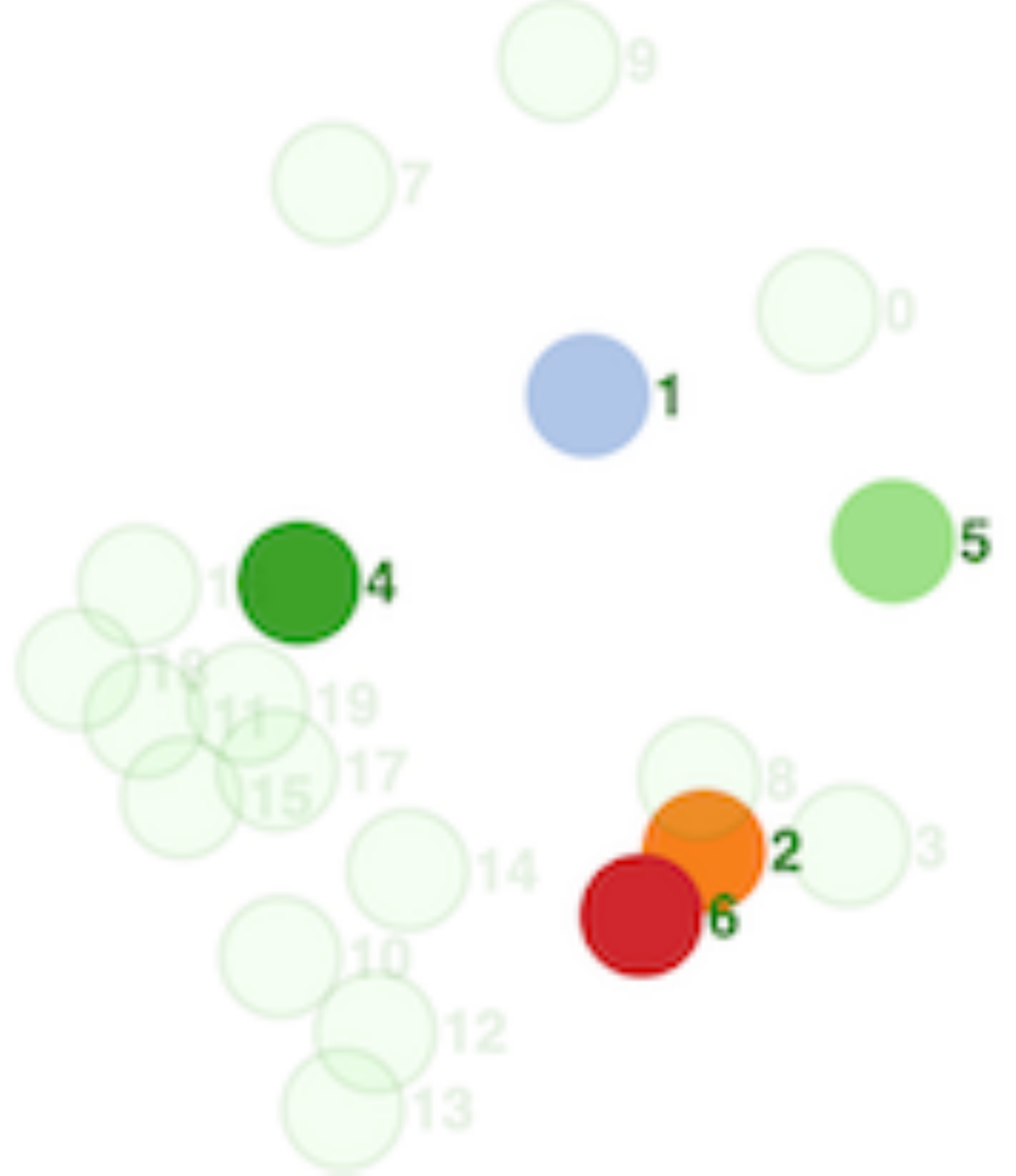}}
    \hlinesep
    \subfigure[]{
    \label{fig:Dire1}%
    \includegraphics[width=0.40\columnwidth]{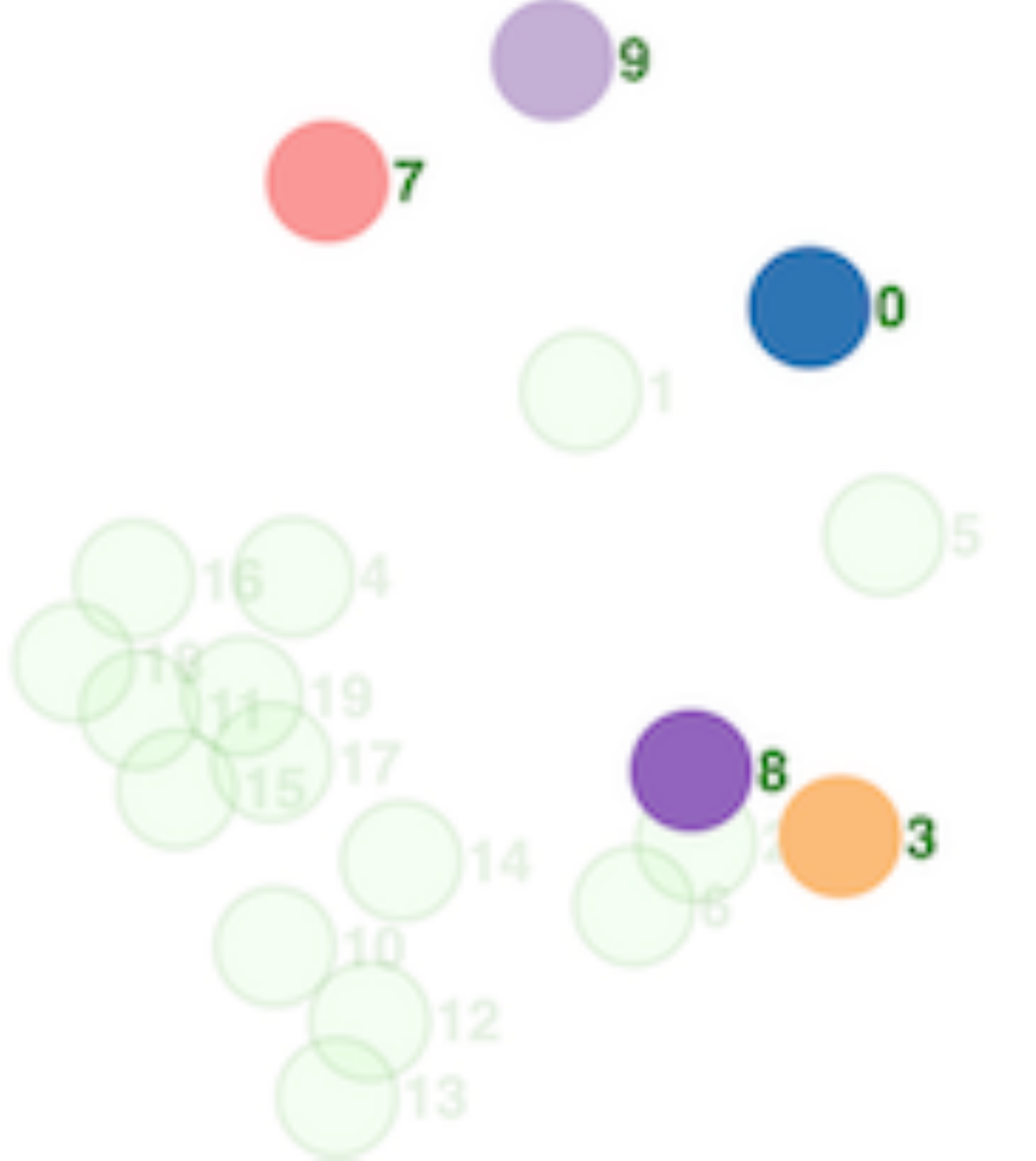}}
    \caption{Sequence graph showing the members of (a) Radiant team and (b) Dire team in the first \textit{DotA 2} match.}
    \label{fig:Dotagame1}
\end{figure}

\begin{figure}
    \centering
    \includegraphics[width=0.7\columnwidth]{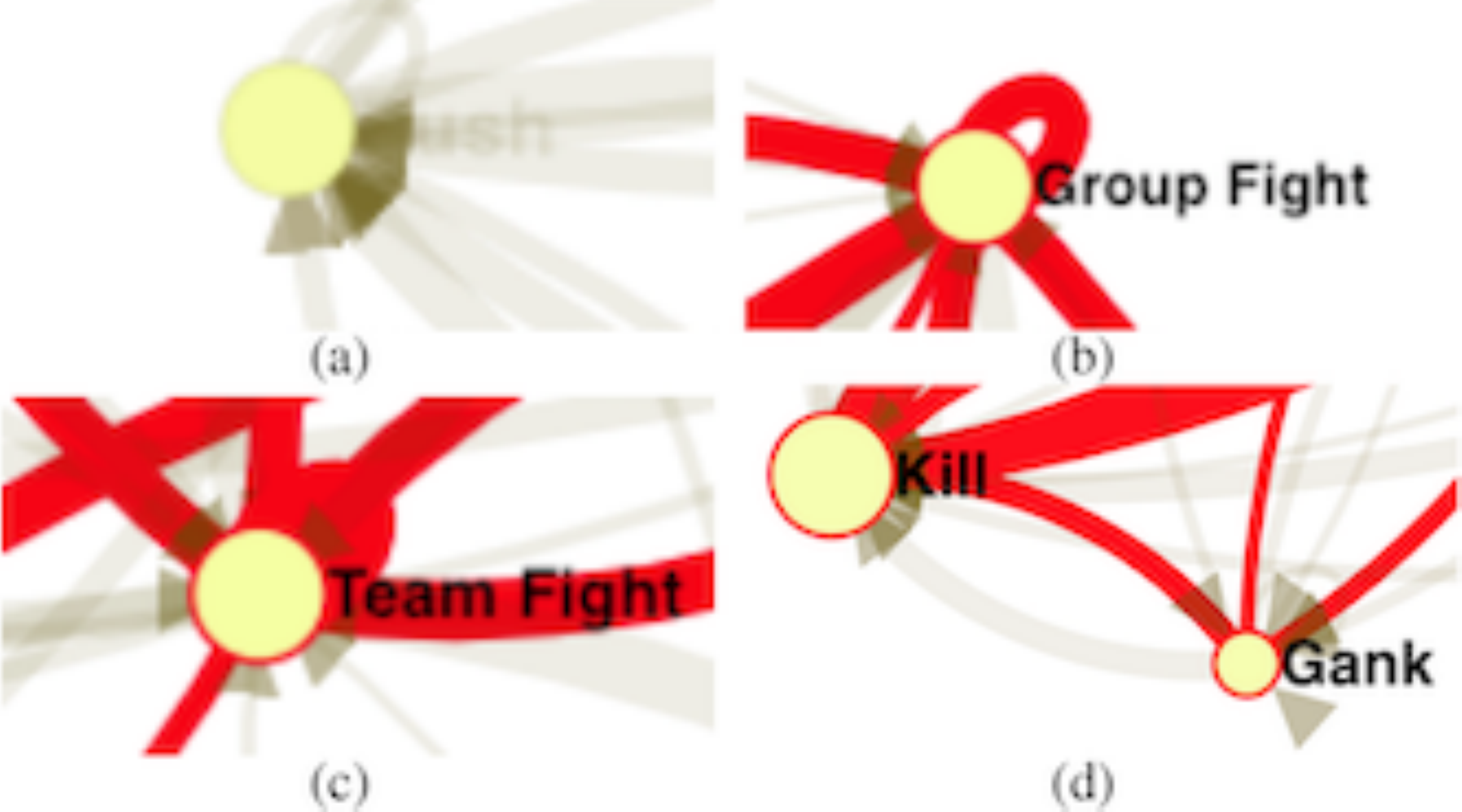}
    \caption{Highlights of the state graph for the hero Batrider, a carry from the first game's Radiant team who played in the top lane, whose gameplay strategy was not characterized by offensive behaviors.}
    \label{fig:BRHighlights}
\end{figure}

\textbf{Data.}
Replay data of \textit{DotA 2} are available on various community websites\footnote{\href{https://www.dotabuff.com}{https://www.dotabuff.com}, \href{https://github.com/skadistats/clarity}{https://github.com/skadistats/clarity}}. A \textit{DotA 2} match replay contains all actions done by players as well as NPCs, and relevant event information such that the game engine can recreate the match as a playback.

\textit{DotA 2} matches are much longer than \textit{BoomTown} levels and include diverse activities and player roles. We randomly selected two matches from professional leagues, with ids: 2593668385 and 3183541562, lasting: 74 and 52 minutes, respectively. The replay data was then visualized and labeled using \textit{StratMapper}. Later, the labeled data was visualized in \textit{Glyph} to analyze the player and team behaviors. It took approximately 10 hours to label these two games. The IRR score using Cohen's Kappa measure was computed for those labels resulting in .72 with 78\% agreement, which was deemed acceptable.

\subsection{Results}\label{subsec:result_dota2}

\subsubsection{Game 1}

Figure~\ref{fig:Dotagame1} shows the sequence graph detailing players' behaviors in the first \textit{DotA 2} game. Players on the Radiant and Dire teams are highlighted on the left and right, respectively. 
In this particular match, the Dire team won the game, and we observed that it was a fairly one-sided match, with the Dire team gaining an early lead and maintaining it throughout the game. In the context of the game, this means that the Dire team was dominating the lanes and had access to more resources, such as jungle encampments, while the Radiant team was forced to play with limited access to such important elements of gameplay, and likely had lower experience points and gold than the Dire side. Figure~\ref{fig:Dotagame1} indicates that players in this game played distinctly differently, as reflected by the distances between individual nodes in the sequence graph. 


Looking more closely at the state graphs of two players from this game, we can see, more specifically, the kinds of differences in gameplay that occurred. Batrider was the top lane hero for the Radiant team. Figure~\ref{fig:BRHighlights} displays highlights of several states from his state graph, where we see that this hero never pushed (no colored links to ``Push'' state in Figure~\ref{fig:BRHighlights}(a)) and never transitioned from a ``Gank'' to a ``Kill'', as seen in Figure~\ref{fig:BRHighlights}(d), indicating no ganks that ended in a kill.

\begin{figure}
    \centering
    \includegraphics[width=0.7\columnwidth]{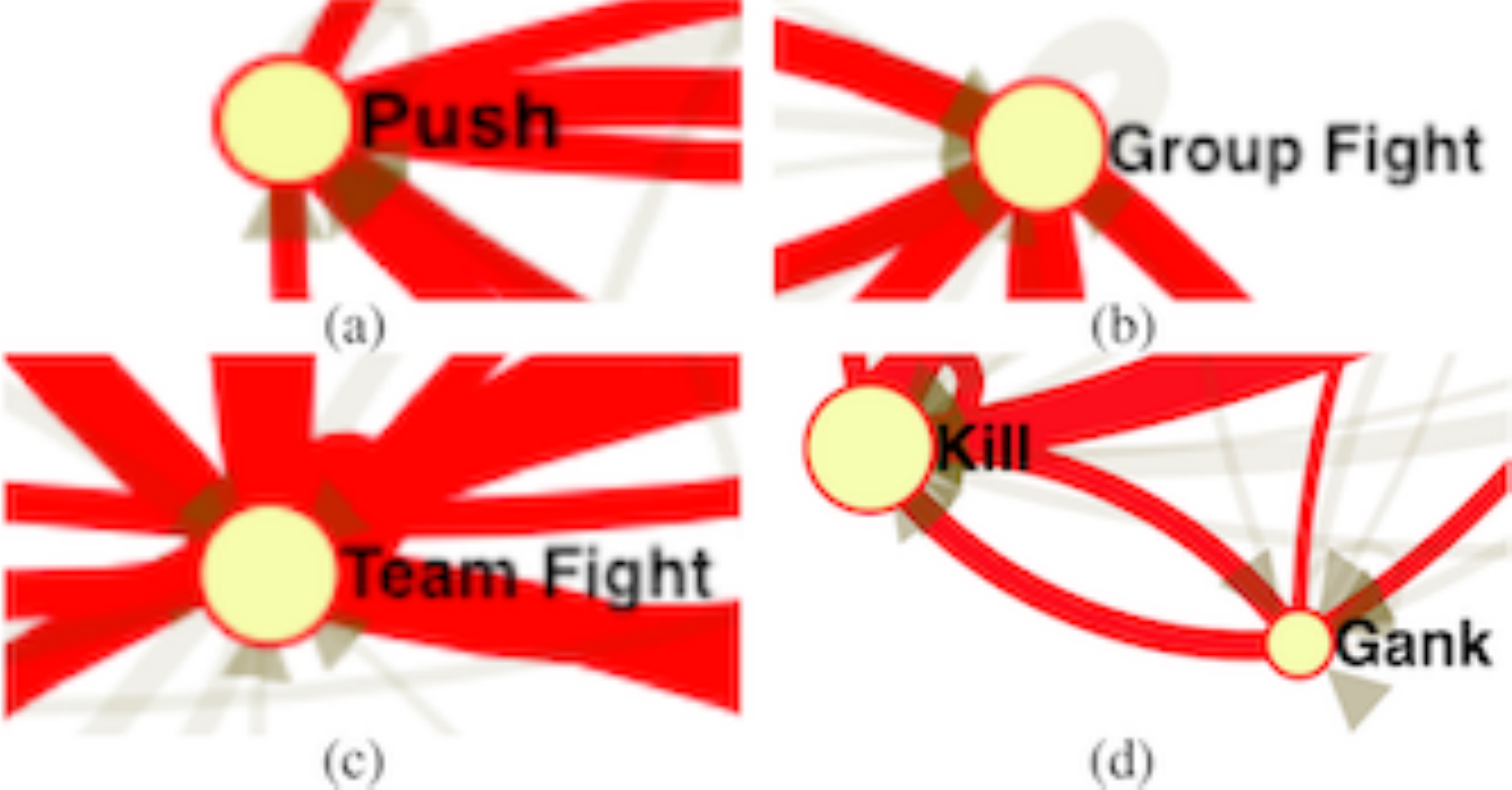}
    \caption{Highlights of the state graph for the hero Slark, a carry from the first game's Dire team, and Batrider's top lane opponent, displaying traces of an offensive strategy indicative of being in an advantageous position.}
    \label{fig:SlarkHighlights}
\end{figure}

\begin{figure}
    \includegraphics[width=\columnwidth]{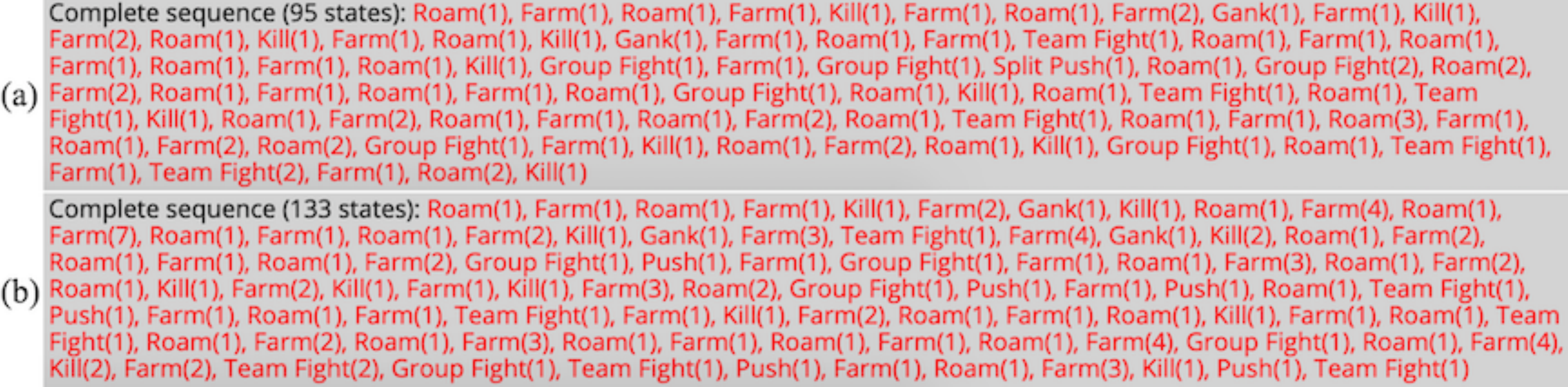}
    \caption{Complete state sequences from the first match for (a) Batrider and (b) Slark, the top lane carries for the Radiant and Dire teams respectively.}
    \label{fig:BatSlarkSeq}
\end{figure}

A more offensive approach is reflected by Slark, the top lane hero for the Dire team, seen in Figure~\ref{fig:SlarkHighlights}. Slark transitioned through the ``Push'' state (Figure~\ref{fig:SlarkHighlights}(a)), indicating that at some point in gameplay he enacted the offensive behaviors of downing enemy structures and killing creeps in order to move the action toward the Radiant side of the field. Additionally, Slark has a transition from ``Gank'' to ``kill'' Figure~\ref{fig:SlarkHighlights}(d), indicating a successful gank that ended in the death of an enemy. This indicates an offensive strategy, focused on relentlessly threatening the opponent with aggressive or offensive behavior, known among the gameplay community as ``applying pressure''. The act of ``applying pressure'' is commonly performed by teams in advantageous positions. Therefore, these pressure related behaviors seen in \textit{Glyph} indicate that the Dire team likely had an advantage throughout the game.

Additionally, towards the end of Batrider's sequence, seen in Figure~\ref{fig:BatSlarkSeq}(a), we see several visits to the ``Roam'' and ``Farm'' states, indicating that Batrider spent much of the end game traveling across the map and killing creeps, rather than attempting to push or participate in team fights. From this, we conclude that his late game behavior was characterized by a need to traverse the map in order to prevent enemy creeps from advancing down any of the lanes. Additionally, it is likely that he was attempting to farm jungle camps to try and catch up to his opponents on gold and experience that he was unable to acquire earlier. These could be characterized as defensive or catch-up-oriented strategic behaviors, and are indicative of a team that has spent much of the game in a losing position, unable to go on the offensive, and desperate to catch up on lost resources.

\begin{figure}%
    \centering
    \subfigure[]{
    \label{fig:boomtown_round6_high}%
    \includegraphics[width=0.40\columnwidth]{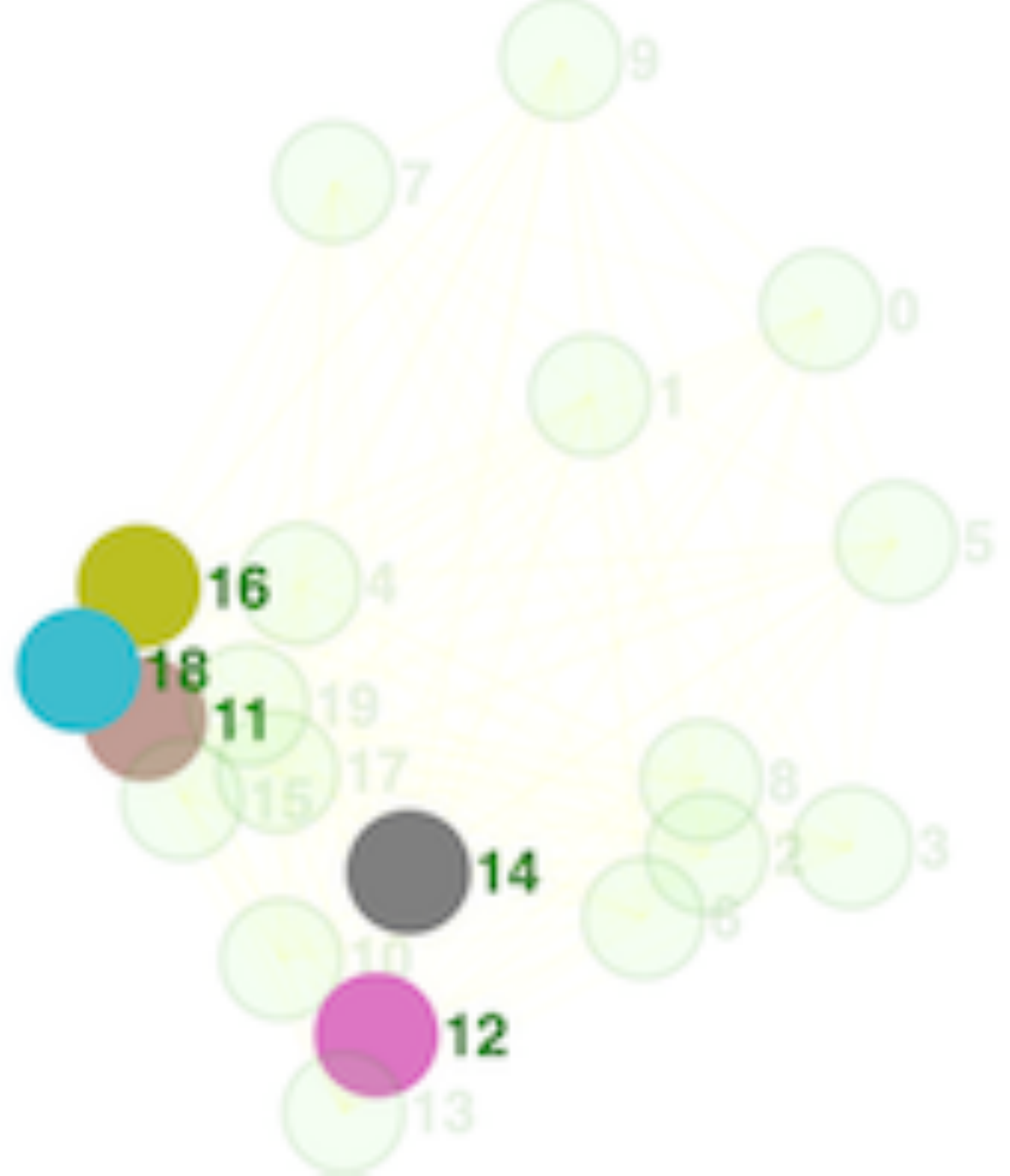}}
    \hlinesep
    \subfigure[]{
    \label{fig:boomtown_round6_low}%
    \includegraphics[width=0.40\columnwidth]{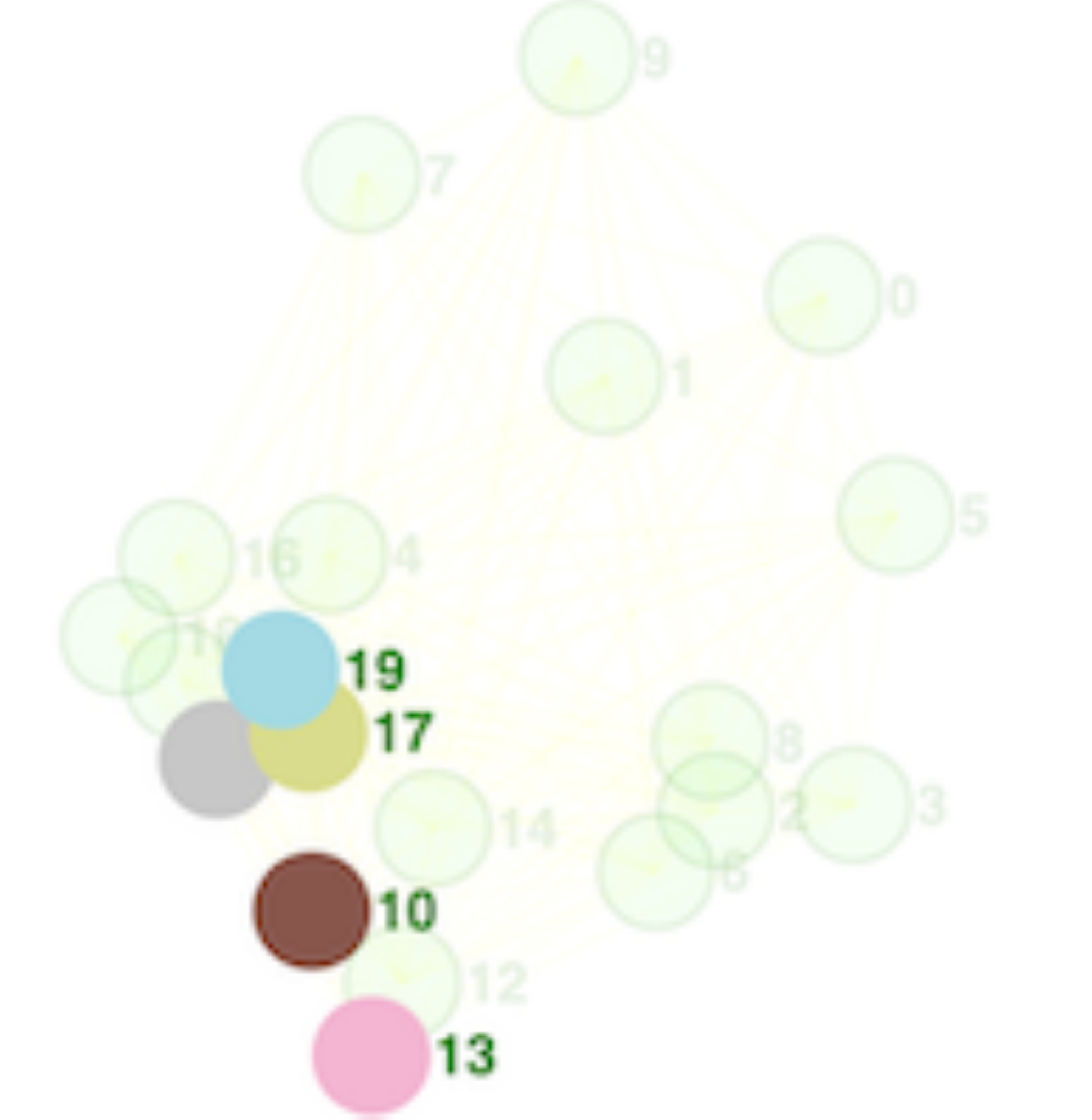}}
    \caption{Sequence graph showing the members of (a) Radiant team and (b) Dire team in the second \textit{DotA 2} match.}
    \label{fig:Dotagame2}
\end{figure}

\begin{figure}%
    \centering
    \subfigure[]{
    \label{fig:Radiant1}%
    \includegraphics[width=0.48\columnwidth]{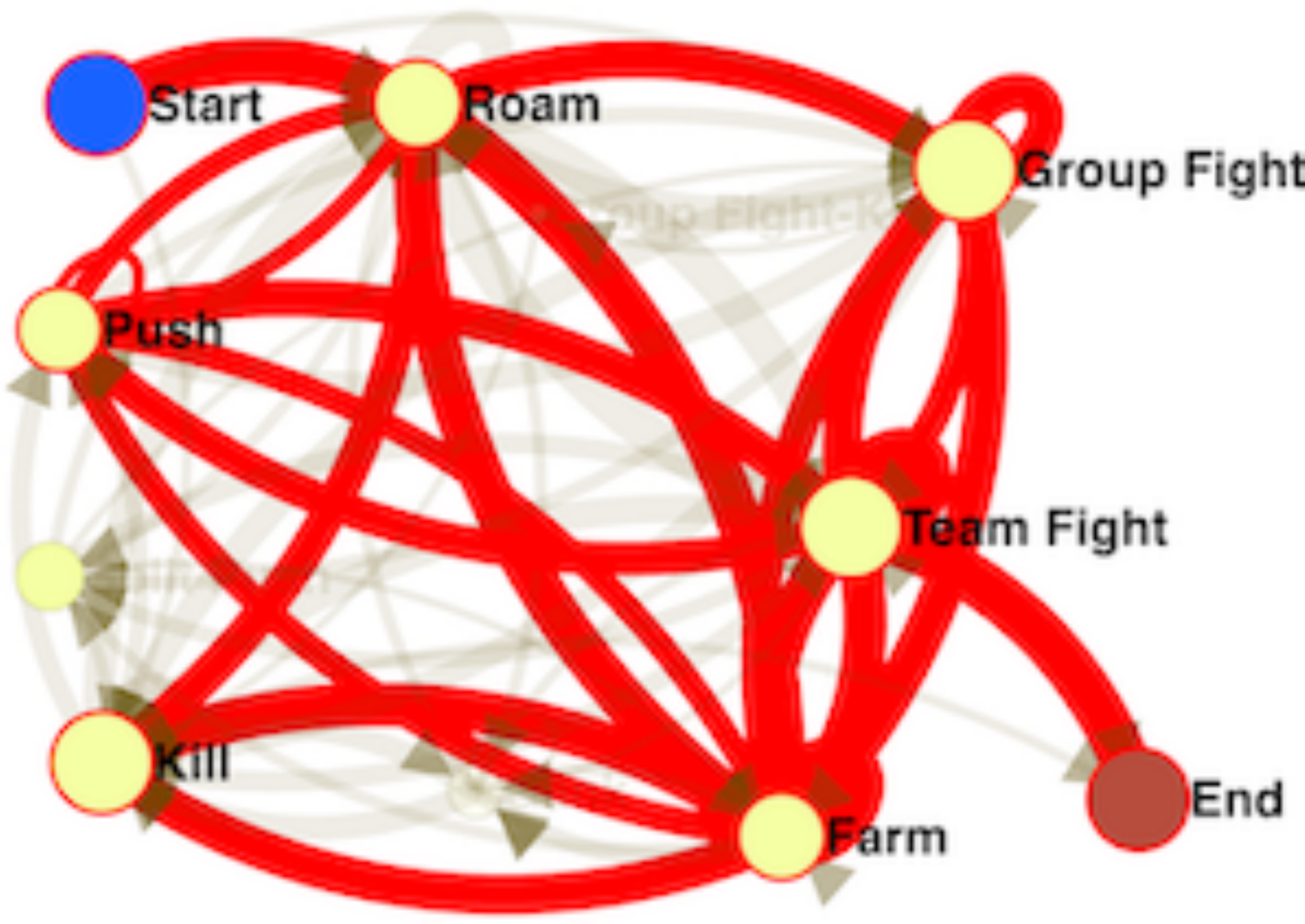}}
    \subfigure[]{
    \label{fig:Dire1}%
    \includegraphics[width=0.48\columnwidth]{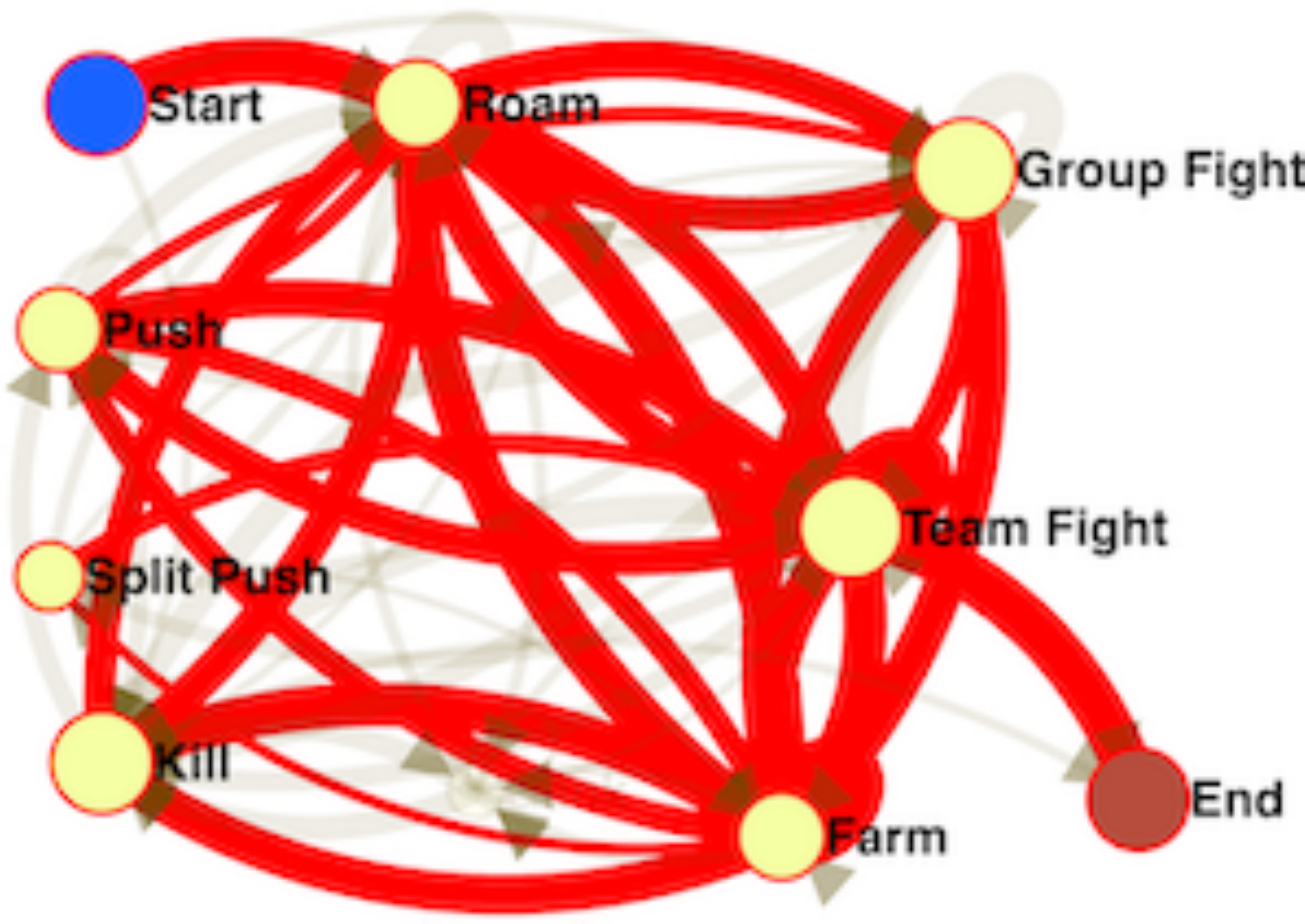}}
    \caption{State graphs of (a) Luna, a carry on the second match's Dire team, and (b) Legion Commander, a carry on the second match's Radiant team. Both sequences are incredibly similar, with both heroes visiting almost the same states.}
    \label{fig:LunavsLC}
\end{figure}

In comparison, the state sequence of Slark is longer and the end of the sequence is characterized more by the presence of ``Team Fight'', ``Group Fight'', and ``Push'' states, as shown in Figure~\ref{fig:SlarkHighlights}(a, b, c) and Figure~\ref{fig:BatSlarkSeq}(b). While ``Roam'' and ``Farm'' are still present in Slark's sequence, they are not the dominant states at the end of the game. From this visualization, we conclude that Slark spent much of the later game states keeping the lanes pushed to the losing side of the map, and engaged in frequent combat in which the winning team, with their gold and experience lead, would have a distinct advantage. These types of behaviors are characteristic of a leading team that is working to secure a win. In contrast, the infrequency of late game ``Group Fight'' or ``Team Fight'' behavior in Batrider's sequence, seen in Figure~\ref{fig:BRHighlights}(b, c) and~\ref{fig:BatSlarkSeq}(a), indicates a hesitance to engage in combat against the opponent, likely due to low chances of victory.

\subsubsection{Game 2}

\begin{figure}
    \includegraphics[width=\columnwidth]{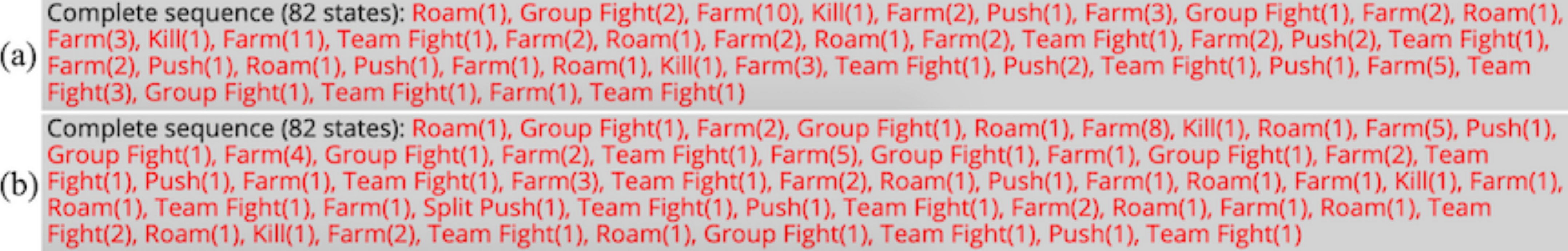}
    \caption{Complete state sequences from the second match for the carries (a) Luna, of the Dire, and (b) Legion Commander, of the Radiant.}
    \label{fig:LunaLegionSeq}
\end{figure}

The second game, whose players are highlighted in Figure~\ref{fig:Dotagame2}, had a distinctly different gameplay scenario. The visualizations illustrate that, unlike the one-sided first game, the second game was very ``back-and-forth'' with both teams experiencing intervals of time in which they had the upper hand. In \textit{Glyph}, we see that the trajectories of all of the players in this game for both teams are very close together (Figure~\ref{fig:Dotagame2}). This is illustrated by the state graphs of two of the carries within the game, seen in Figure~\ref{fig:LunavsLC}. Luna, of the Dire team, and Legion Commander, of the Radiant team, possess extremely similar state graphs. Both visited almost the same set of states, with the only differences being the order, transitions, and minute variances in frequency of individual states (both visit the same number of states) as can be seen in Figure~\ref{fig:LunaLegionSeq}.

\begin{figure}
    \centering
    \includegraphics[width=0.3\columnwidth]{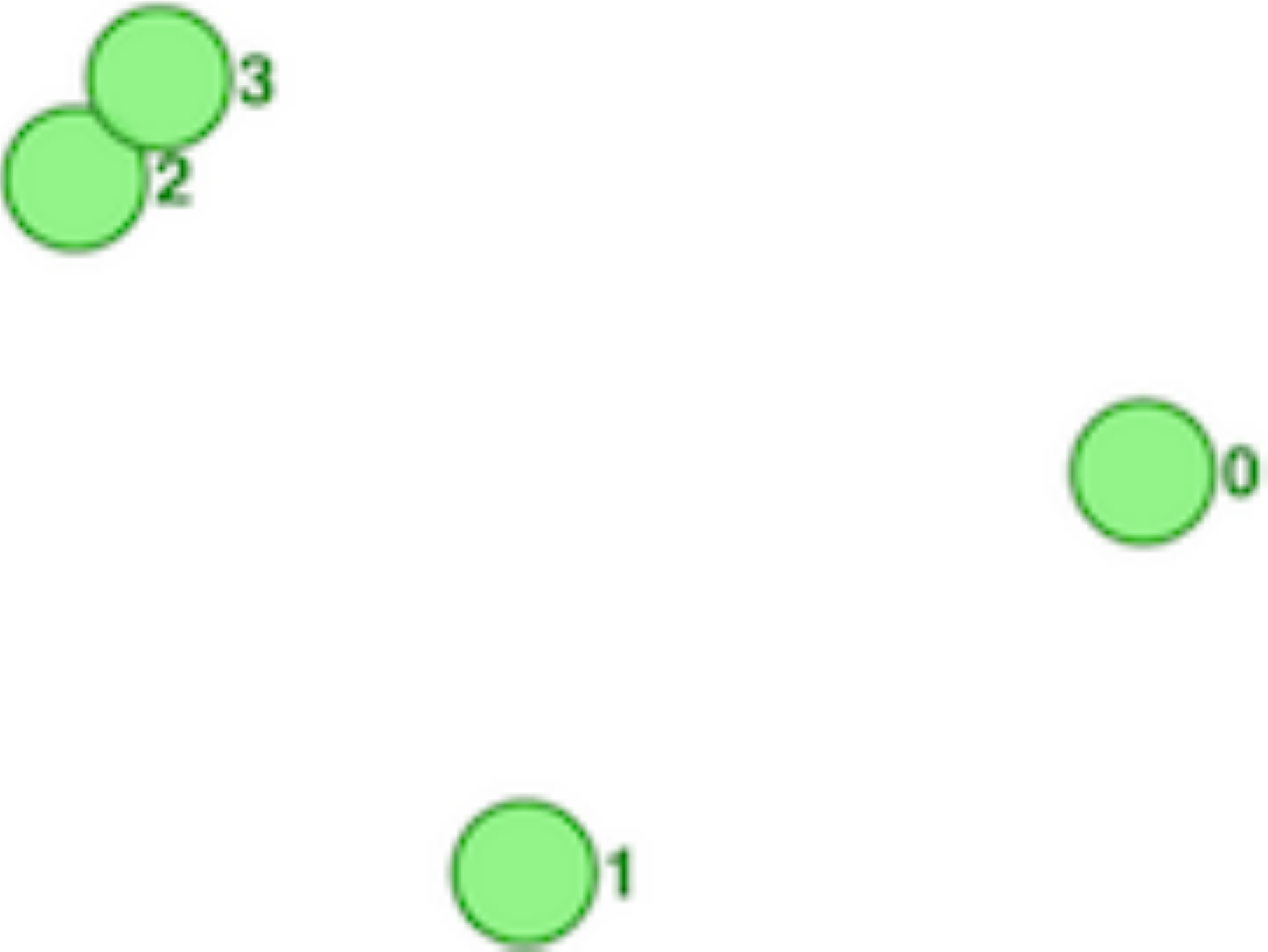}
    \caption{Nodes representing team trajectories across the two \textit{DotA 2} games as shown in the group graph. Teams 0 (Dire) and 1 (Radiant) played each other in the first (one-sided) match and teams 2 (Dire) and 3 (Radiant) played each other in the second (back and forth) match.}
    \label{fig:traj}
\end{figure}

In both Luna and Legion Commander's sequences, Figure~\ref{fig:LunaLegionSeq}, we observed the offensive behaviors of pushing and fighting (discussed as being prominent in the sequence of Slark from the first game), as well as the defensive behaviors of late game roaming and farming (prominent in the sequence of Batrider). Thus, it is possible to conclude that, at various points throughout gameplay, the Radiant and Dire teams in the second game experienced both winning and losing states, and that they behaved accordingly, indicating a much closer match than the one-sided first game. This observation is supported by the visualization of the team trajectories, where teams `0' (Dire) and `1' (Radiant) from the first match had distinctly different sequences, while teams `2' (Dire) and `3' (Radiant) from the second match had very similar ones as shown in Figure~\ref{fig:traj} by the distance between the nodes. These results demonstrate how \textit{Glyph} can be used to identify player behaviors and in turn infer various game states that likely influenced those behaviors.

\subsubsection{Role Based Gameplay}

\begin{figure}%
    \centering
    \subfigure[Radiant team: 2, 6 and Dire team: 3, 8.]{
    \label{fig:supports}
    \includegraphics[width=0.40\columnwidth]{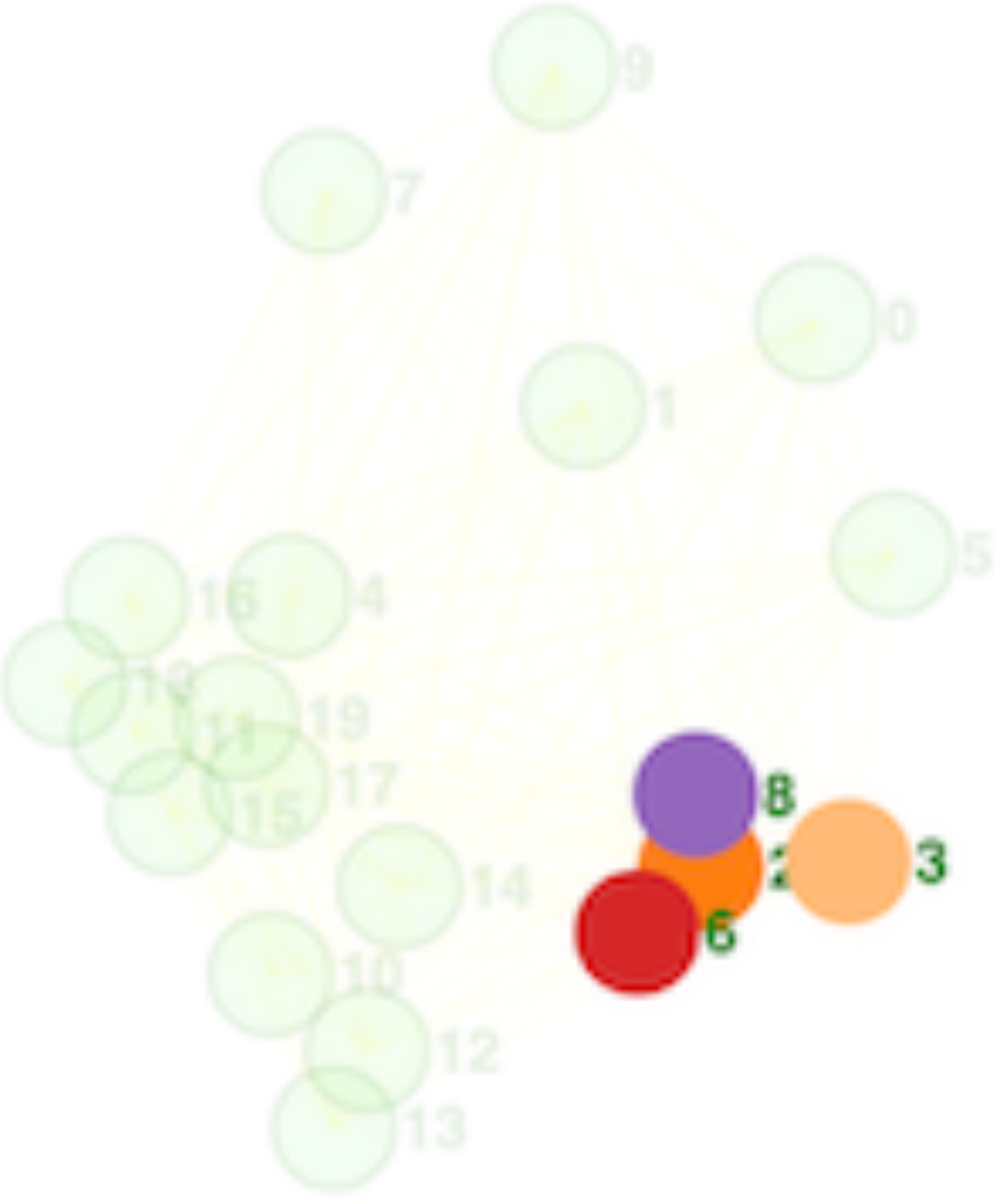}}
    \hlinesep
    \subfigure[Radiant team: 12, 14 and Dire team: 10, 13.]{
    \label{fig:supports2}%
    \includegraphics[width=0.40\columnwidth]{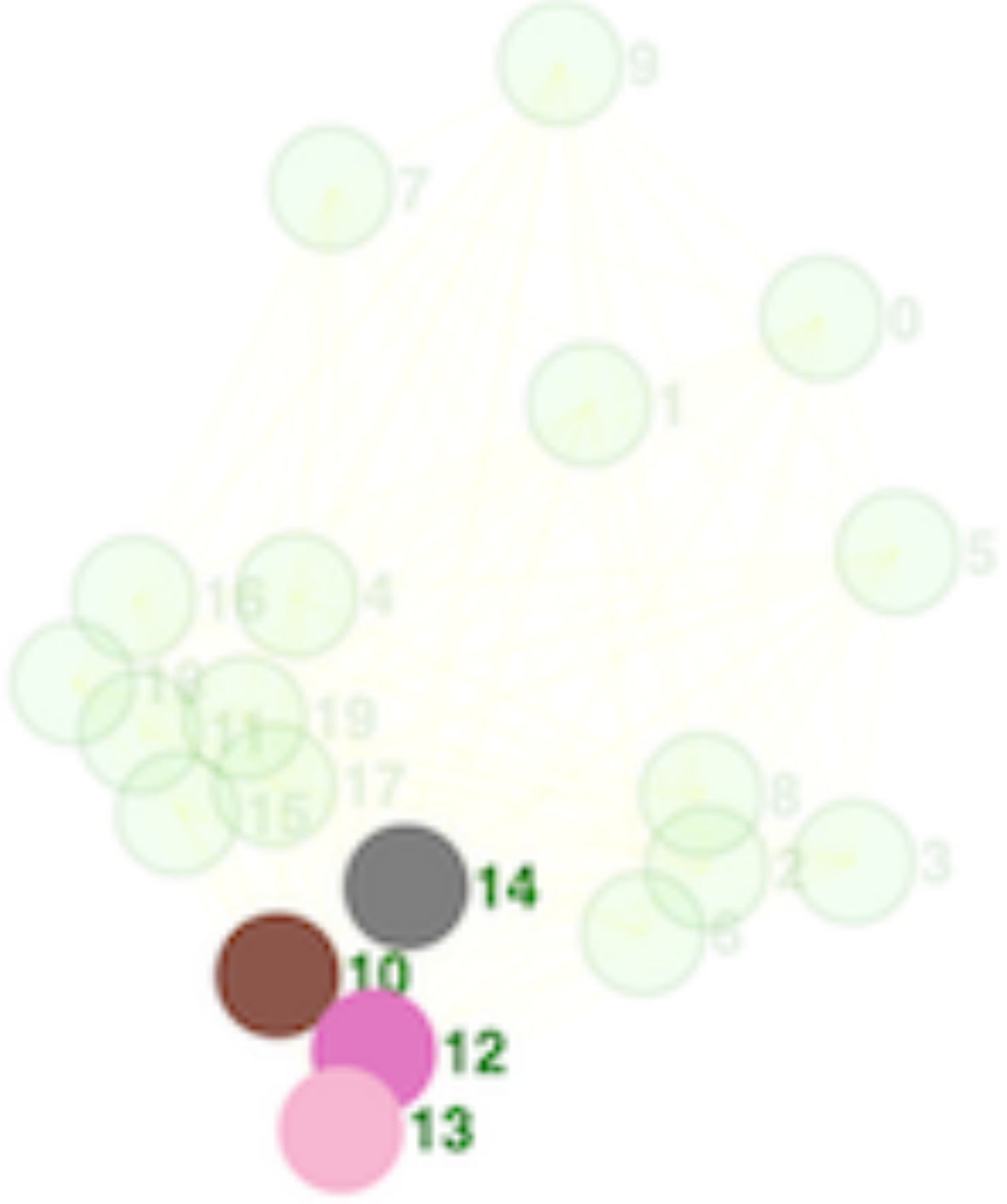}}
    \caption{Similar sequences of players playing support roles in: (a) first match, (b) second match.}
    \label{fig:supports_two_match}
\end{figure}

We also observed differences in strategic behavior based on role. In the sequence graphs shown in Figure~\ref{fig:supports_two_match}, we see that, in both matches, there were players across teams who played similarly to each other. In the first match, four players across teams (Figure~\ref{fig:supports_two_match}(a)), specifically players `2' and `6' of the Radiant team, and players `8' and `3' of the Dire team, all had similar sequences to one another; these players were all playing heroes who can be categorized as supporting role heroes in \textit{DotA 2}. While the core heroes of each team possessed sequences distinctly different from each other (Figure~\ref{fig:Dotagame1}), the support heroes' sequences were similar to each other both within and across teams, indicating that the role and strategy of a support player is likely less affected by the fact that the team was in the lead or not, and thus, support players will play similarly regardless of these factors. 



\begin{figure}%
    \centering
    \subfigure[]{
    \label{fig:boomtown_round6_high}%
    \includegraphics[width=0.48\columnwidth]{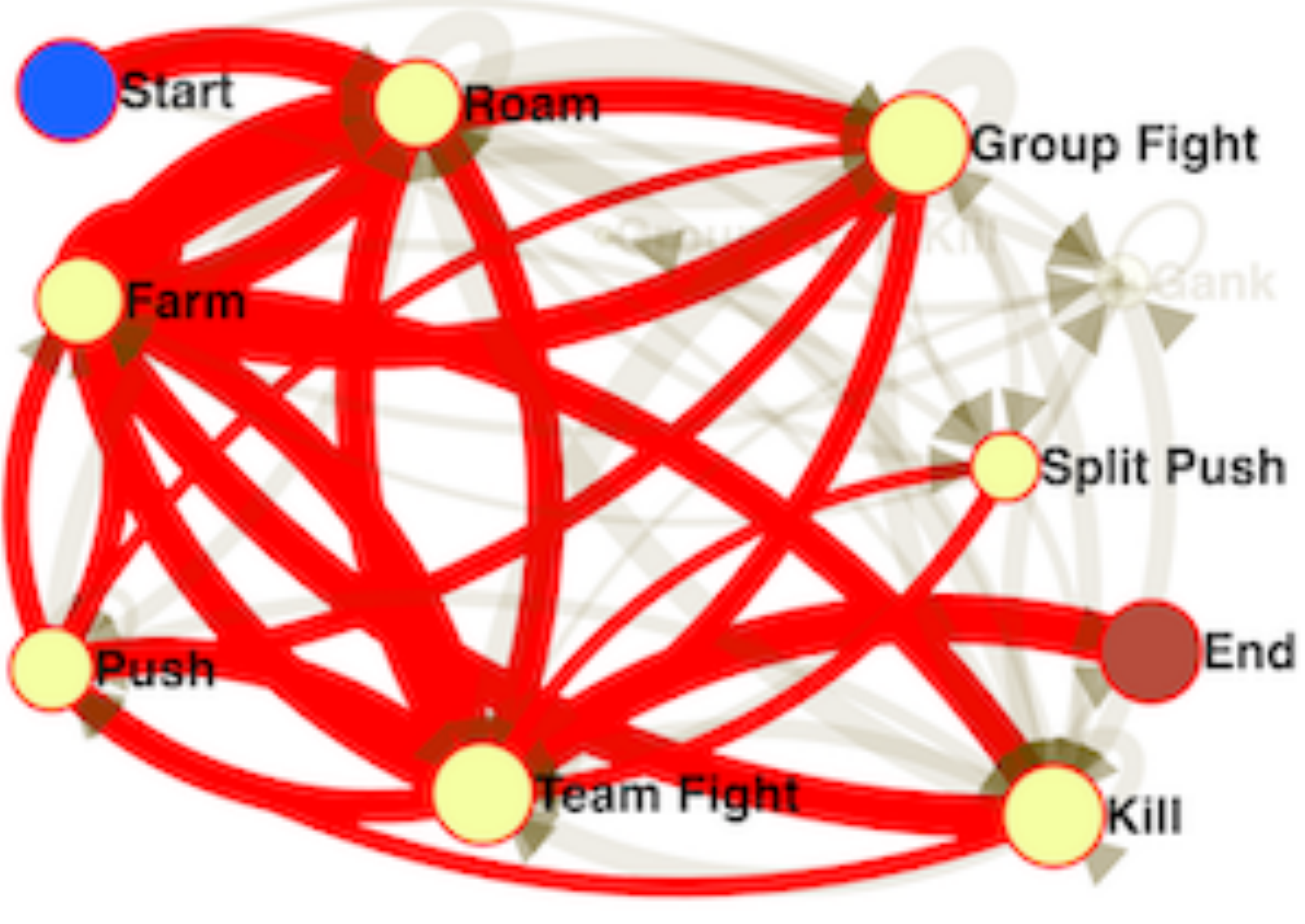}}
    \subfigure[]{
    \label{fig:boomtown_round6_low}%
    \includegraphics[width=0.48\columnwidth]{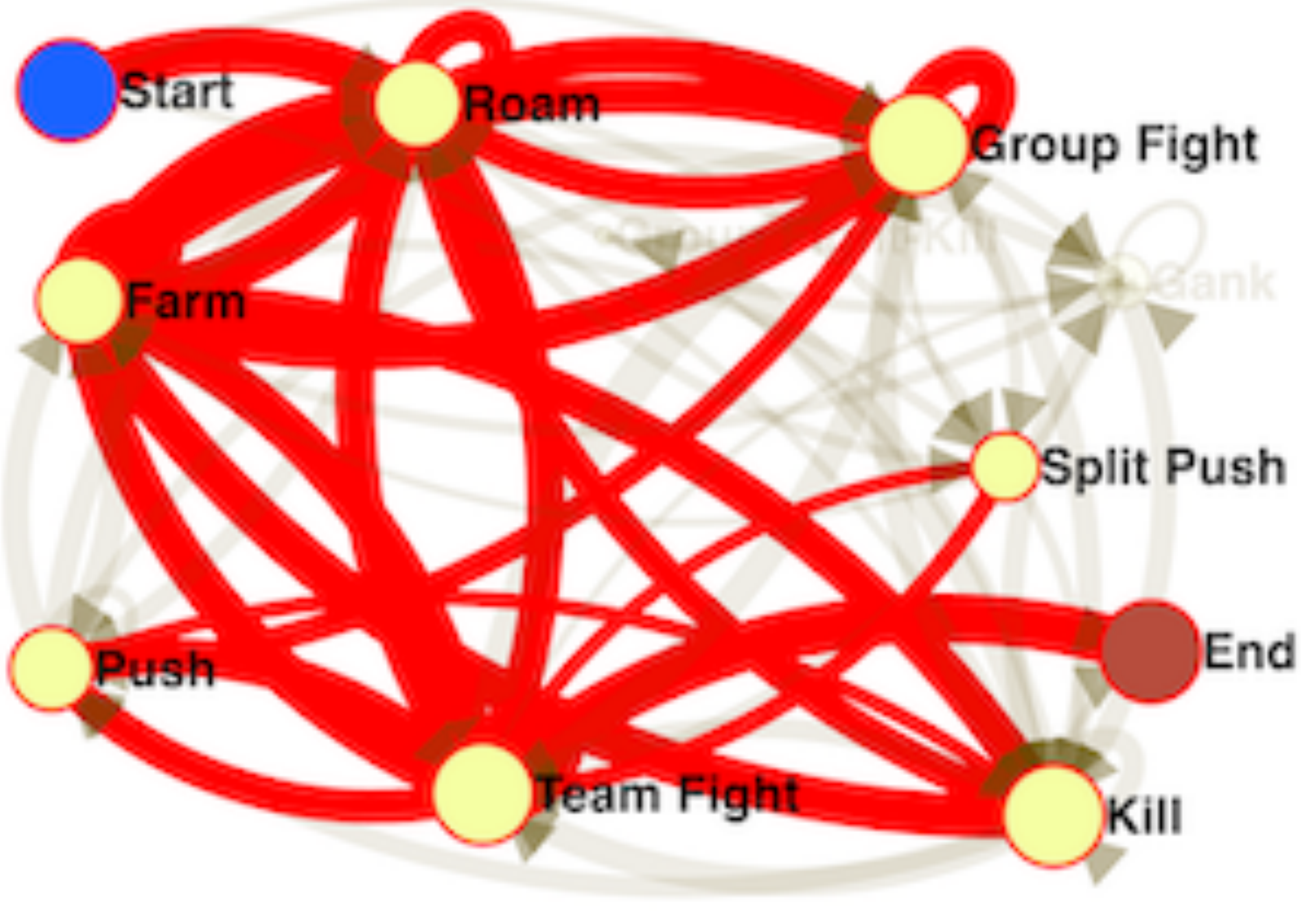}}
    \caption{State graphs for (a) Shredder, a carry, and (b) Crystal Maiden, a support, both for the Radiant team in the second match.}
    \label{fig:CrystalvsShred}
\end{figure}

\begin{figure}
    \includegraphics[width=\columnwidth]{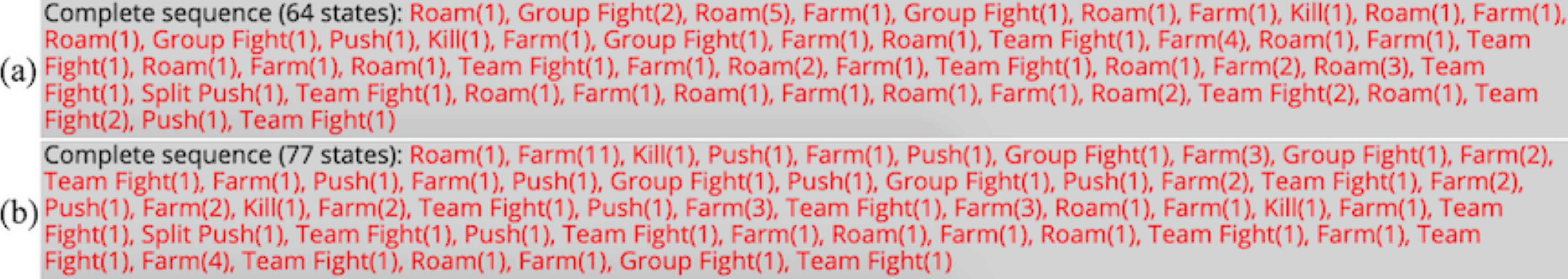}
    \caption{Complete state sequences from the second match for (a) Shredder, a carry, and (b) Crystal Maiden, a support. Both heroes were on the Radiant team}
    \label{fig:CrystShredSeq}
\end{figure}

Similarly, in the second game, heroes played in supporting roles have similar sequences, highlighted in Figure~\ref{fig:supports_two_match}(b), that can be observed to be distinctly different than those of the teams' core heroes, highlighted in Figure~\ref{fig:Dotagame2}. Players on \textit{DotA 2} teams are expected to adopt and perform roles, defined by the game and its design, and take actions based on the expectations of those roles. We, therefore, are able to clearly see using \textit{Glyph} this role-based gameplay and how players show distinct transitions through states that reflect these roles with support players in both matches displaying different sequences than their respective core players, regardless of the gameplay conditions. This is illustrated further in Figure~\ref{fig:CrystalvsShred} where Shredder, a carry on the second game's Radiant team can be seen to visit the ``Farm'' state quite a bit, as well as the ``Kill'', ``Push'', and various fight states. Crystal Maiden, a support on the same team, spends significantly less time farming and instead frequently visits the ``Roam'' state, as many supports in \textit{DotA 2} roam the map looking for opportunities to assist the carries (Figure~\ref{fig:CrystShredSeq}).

\begin{figure}%
    \centering
    \subfigure[]{
    \label{fig:boomtown_round6_high}%
    \includegraphics[width=0.48\columnwidth]{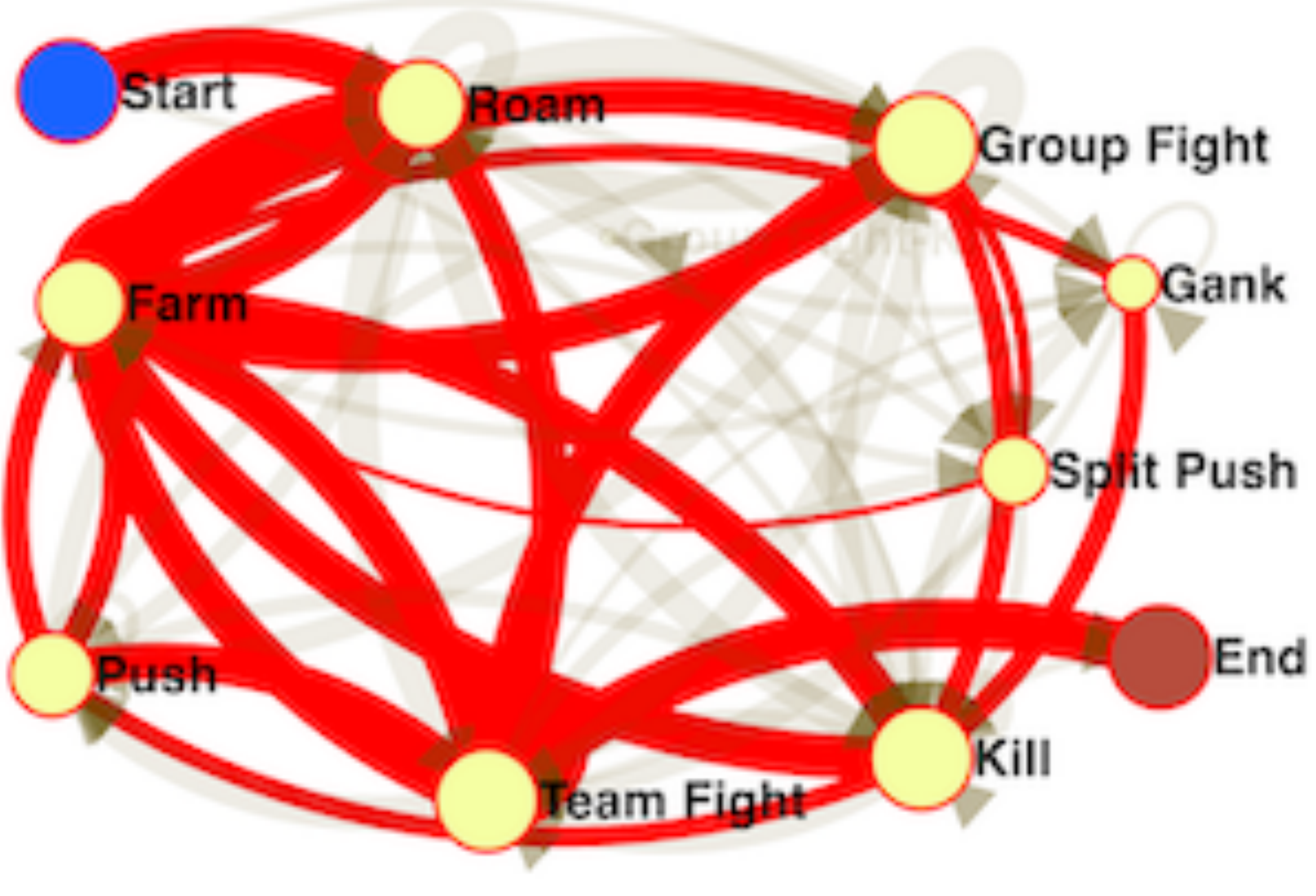}}
    \subfigure[]{
    \label{fig:boomtown_round6_low}%
    \includegraphics[width=0.48\columnwidth]{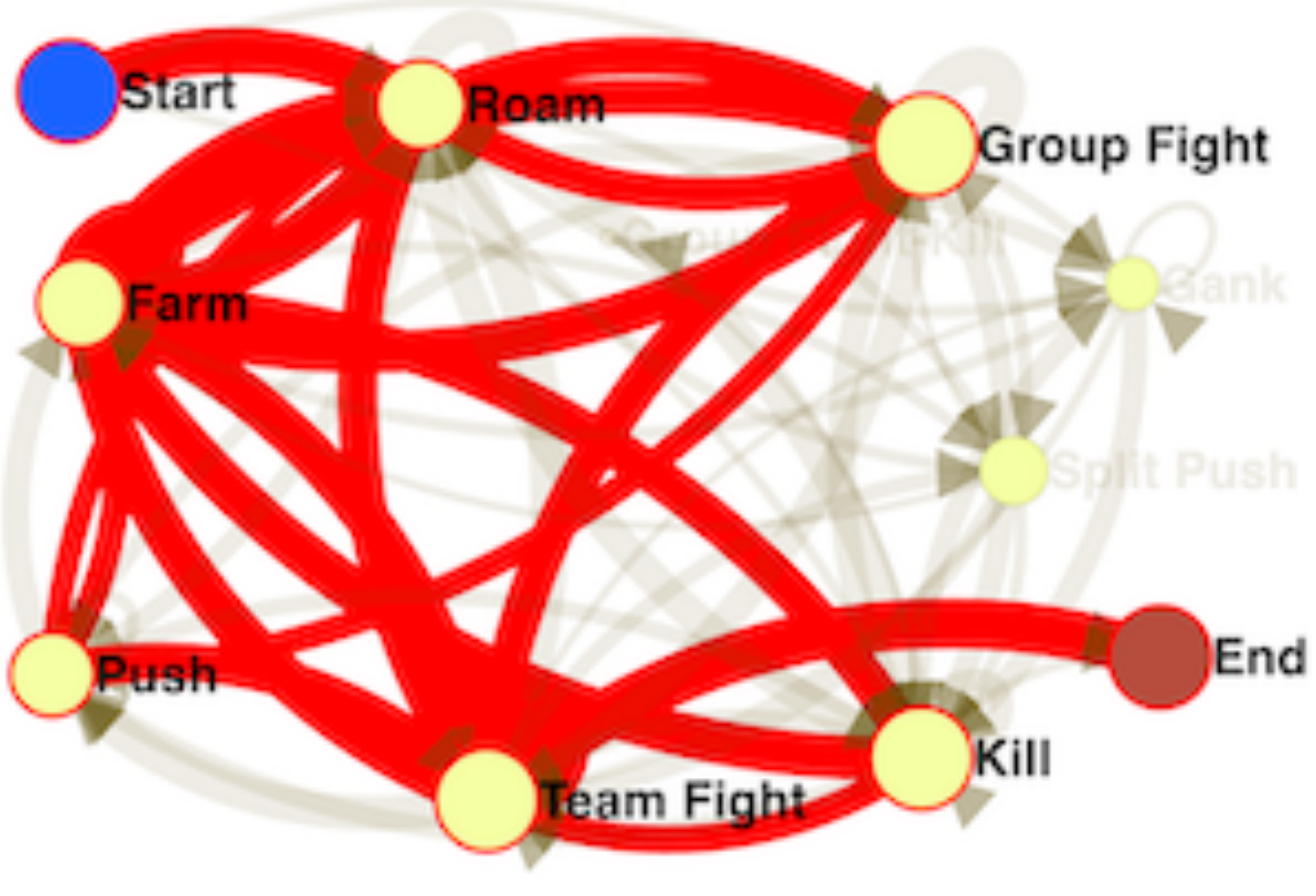}}
    \caption{State graphs for (a) Alchemist, the mid lane carry for the Radiant team in the first match, and (b) Darkseer, the bottom lane carry for the Radiant team in the second match.}
    \label{fig:darkseervsalch}
\end{figure}

\begin{figure}
    \includegraphics[width=\columnwidth]{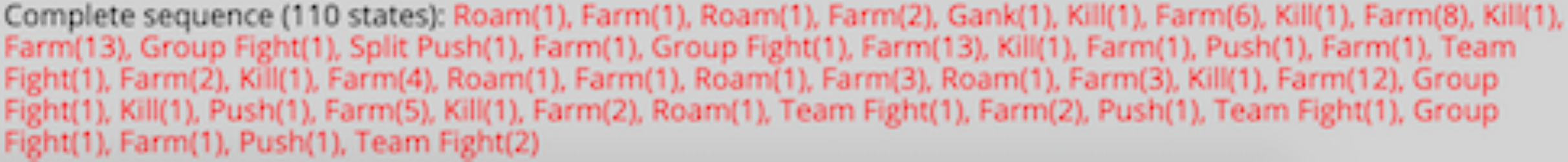}
    \caption{Alchemist's (mid-lane carry on the losing Radiant team) complete sequence from the first match.}
    \label{fig:AlchSeq}
\end{figure}

\subsubsection{Context Specific Strategy}

A final point worth noting is that player `4', a player in the first game (Figure~\ref{fig:Dotagame1}(a)), has a sequence very similar to the players of the second game. This player was playing a hero known as Alchemist. State graphs in Figure~\ref{fig:darkseervsalch} for Alchemist and a carry from the second game, Darkseer, show that the two heroes possessed very similar sequences of states. Specifically, both heroes spent a large amount of time in the farming state, with Alchemist having a very large number of transitions from farming to farming. A carry's job in a game of \textit{DotA 2}, according to the gameplay community\footnote{\href{https://liquipedia.net/dota2/Farm_Dependency}{https://liquipedia.net/dota2/Farm\_Dependency}, \href{https://gosu.ai/blog/dota2/positions-and-roles-guide/}{https://gosu.ai/blog/dota2/positions-and-roles-guide}} is to focus on farming in order to gain gold and experience, so that they may become powerful during late game and ``carry'' the team to victory. However, when a team is in a losing state, it is difficult to do so, due to lack of resources and a need to focus on defense. Thus, the sequences of the carries on the losing team, like Batrider, tend to display signs of defensive end game behavior like farming and roaming as shown previously in Figure~\ref{fig:BatSlarkSeq}(a). Alchemist, however, was the exception to this rule. He continued to focus on farming through early and mid game, finding ways to do so and letting his teammates play the defensive roles, and was able to transition into a more offensive role late in the game, as illustrated by his complete state sequence in Figure~\ref{fig:AlchSeq}, which features ``Push'', ``Team Fight'', and ``Group Fight'' in higher frequencies towards the end. Thus, as opposed to the first game, where one team played entirely in a defensive manner, unable to farm carries into late game power, and the other in an offensive manner, with carries able to forgo excessive farming in favor of pressure, Alchemist's sequence places him closer to the sequences of the carries in the second game, where both teams experienced both conditions. 



This reflects a common gameplay strategy employed by losing teams in \textit{DotA 2}, as recognized by the community\footnote{\href{https://www.quora.com/What-are-some-of-the-team-picking-strategies-in-dota}{https://www.quora.com/What-are-some-of-the-team-picking-strategies-in-dota}, \href{https://steamcommunity.com/app/570/discussions/0/618458030687911877}{https://steamcommunity.com/app/ 570/discussions /0/618458030687911877}}, to attempt to stall the game as long as possible while the team's carry focuses on farming so that they may become strong enough to turn the game around. From \textit{Glyph}, it can be hypothesized that Alchemist's team employed this strategy, and hoped to play defensively long enough for Alchemist to catch up to opposing carries in strength, thus choosing to sacrifice their own ability to play offensively in favor of prolonging the game. \textit{Glyph} not only illustrates the exact, granular nature of this well known gameplay strategy, but also allows us to see, hypothesize, and draw conclusions regarding the ways in which the game state may be affecting players' abilities or decisions to pursue such strategies and tactics. 

\section{Discussion}

There are several advantages to using \textit{IBA} as a methodology over state of the art methods, where aggregated data, clustering, or classification methods are applied to game data. For example, while previous work utilized clustering to identify behaviors based on character roles, \textit{IBA} provides a deeper understanding of such behaviors. It allows for the visualization of not only the actions taken, but also the sequence and timing of such actions, thus giving researchers a deeper contextualization of actions within the scope of the rest of the game. This allows researchers to infer the reason certain actions were taken from a strategic standpoint. Further, our labels are defined based on existing standards of the gameplay community, and therefore reflect the types of strategic considerations and behaviors that players of the game would take. This context-based label derivation allows \textit{Glyph} to reveal more about the situational nature of the game being visualized, such as who was in the lead and how players responded, as seen in game 1 of \textit{DotA 2}, discussed in the first part of the results section. As strategy in many games is pursued in response to environmental factors within the game, such a context-rich visualization is key to discovering insight from data. Additionally, the case studies that apply this approach to both \textit{BoomTown} and \textit{DotA 2} show the generalizability of \textit{IBA} as a methodology. 


The labeling process is labor extensive, and the time taken to apply each label depends on the complexity, frequency, and duration of the behavior. However, having a human-in-the-loop strengthens the context aware nature of the visualization. By having humans develop the labels, our methodology engages a process of humans interpreting the actions of other humans. Thus, \textit{Glyph} represents data as a sequence of actions defined by a human within the context of the game, increasing the chances of extracting understandable and meaningful behavioral information. Further, having a human involved in the interpretation process, especially one who was also involved in the labeling process, allows the data to be interpreted with its context in mind. Given the importance of contextual and situational considerations when interpreting action sequence data, having a human who is familiar with the context do the interpretation is critical to obtaining reliable results. Nevertheless, as we will discuss in future work, we are engaging in further work to incorporate automatic detection given human authored behaviors to aid in the process of labeling.

\section{Conclusion}
\label{sec:conclusion}

In this paper, we proposed \textit{IBA}, a methodology, comprised of two visualization systems and a labeling mechanism that allows researchers and developers to label player and team behaviors and then visualize them in order to understand the underlying tactics, complex dynamics, and strategic decisions made within given game contexts. \textit{IBA} addresses the issue of generalizability of previous work and enables researchers and designers to understand player and team tactics and strategies in an interpretable way. This allows game designers and developers to better identify and understand imbalances in game mechanics during the development cycle.


We have identified multiple directions for future work. As the labeling process is a time-consuming and labor extensive procedure, mathematical models, such as~\cite{sapienza2018non, cavadenti2016did, miller2010group}, can be used as a basis for the labeling process; i.e. give labeling experts some suggested labels to explore or edit. Related to this, we are currently actively investigating the possibility of developing an inference system to label unlabeled data using the labeled data. Specifically, we are investigating the use of deep learning and other classification techniques for that purpose. Additionally, we are considering automatic event analysis tools~\cite{chu2017broadcasted, ringer2018deep, intharah2018deeplogger} that can detect events from game video, which can be used for automatic detection of behaviors and events. However, it should be noted that the current state of the art methods are limited, and cannot be used at this time due to the simple nature of events detected and the inability of these methods to deal with complex high level behaviors like the ones discussed in this paper. Another interesting direction of future research is to extend the \textit{StratMapper} interface to include events such as chats, pings, and annotations, because of the importance of non-verbal communications on player performance, workload, and experience in distributed multiplayer games, as well as how they are used for collaborative planning~\cite{wuertz2017players, toups2014framework, alharthi2018investigating, leavitt2016ping, wuertz2018design}.


\section{ACKNOWLEDGEMENTS}
We want to thank everyone who participated in our studies. We also want to thank Gallup for giving us data for their game \textit{BoomTown}. This research was developed with funding from the Defense Advanced Research Projects Agency (DARPA). The views, opinions and/or findings expressed are those of the author and should not be interpreted as representing the official views or policies of the Department of Defense, DARPA, or the U.S. Government. 

\balance{}

\bibliographystyle{SIGCHI-Reference-Format}
\bibliography{ref}

\end{document}